\documentclass{sig-alternate-10pt}
\usepackage{amsmath,amssymb,graphicx,subfigure}
\usepackage{tikz}
\usepackage[pdfborder={0 0 0}]{hyperref}
\usepackage{mathrsfs}

\usepackage{dsfont}

\usepackage{MnSymbol}

\newtheorem{proposition}{Proposition}
\newtheorem{theorem}{Theorem}
\newtheorem{lemma}{Lemma}

\newtheorem{corollary}{Corollary}

\newtheorem{example}{Example}

\title{Regular path queries on graphs with data: \\
A rigid approach\thanks{Supported by the National Natural Science Foundation of China under Grant No. 61100062.}} 

\author{
Zhilin Wu \\
State Key Laboratory of Computer Science, \\
Institute of Software, Chinese Academy of Sciences \\
wuzl@ios.ac.cn
}

\newcommand{\Aa}{\mathcal{A}}
\newcommand{\Bb}{\mathcal{B}}

\newcommand{\Dd}{\mathbb{D}}
\newcommand{\Cc}{\mathcal{C}}
\newcommand{\gdb}{\mathcal{G}}
\newcommand{\dg}{\mathscr{G}}

\newcommand{\Ll}{\mathscr{L}}

\newcommand{\Tt}{\mathcal{T}}

\newcommand{\cur}{{cur}}

\newcommand{\first}{{first}}

\newcommand{\last}{{last}}

\newcommand{\suc}{{suc}}

\newcommand{\pred}{{pred}}

\newcommand{\ren}{{\rm{ren}}}

\begin{document}

\maketitle

\begin{abstract}
Regular path queries (RPQ) is a classical navigational query formalism for graph databases to specify constraints on labeled paths. Recently, RPQs have been extended by Libkin and Vrgo$\rm \check{c}$ to incorporate data value comparisons among different nodes on paths, called regular path queries with data (RDPQ). It has been shown that the evaluation problem of RDPQs is PSPACE-complete and NLOGSPACE-complete in data complexity. On the other hand, the containment problem of RDPQs is in general undecidable. In this paper, we propose a novel approach to extend regular path queries with data value comparisons, called rigid regular path queries with data (RRDPQ). The main ingredient of this approach is an automata model called nondeterministic rigid register automata (NRRA), in which the data value comparisons are \emph{rigid}, in the sense that if the data value in the current position $x$ is compared to  a data value in some other position $y$, then by only using the labels (but not data values), the position $y$ can be uniquely determined from $x$. We show that NRRAs are robust in the sense that nondeterministic, deterministic and two-way variant of NRRAs, as well as an extension of regular expressions, are all of the same expressivity. We then argue that the expressive power of RDPQs are reasonable by demonstrating that for every graph database, there is a localized transformation of the graph database so that every RDPQ in the original graph database can be turned into an equivalent RRDPQ over the transformed one. Finally, we investigate the computational properties of RRDPQs and conjunctive RRDPQs (CRRDPQ). In particular, we show that the containment of CRRDPQs (and RRDPQs) can be decided in 2EXPSPACE.
\end{abstract}

\vspace*{-2mm}
\section{Introduction}

Graph data management is a classical research field in database community and has achieved a recent resurgence, with the momentums from new application domains, such as online social networks, bioinformatics, and semantic web. Various query languages have been proposed for graph databases (see \cite{AG08,Wood12, Bar13} for surveys). Among them, regular path queries (RPQ) are basic query formalisms to specify path constraints in graph databases.

Graph databases are usually modelled as edge-labeled graphs. A RPQ looks for a pair of nodes connected by a path whose sequence of labels belongs to a regular language (\cite{CMW87}). For the convenience of specifications, RPQs can be extended in a natural way, called RPQs with inverse word symbols (2RPQ), to allow traversing edges in both directions. Since the availability of inverse symbols much eases the specifications,  we will focus on path queries with inverse symbols through this paper. 

Extensions of RPQs to specify the relationships among multiple paths have been investigated intensively, e.g. conjunctive RPQs (CRPQ) which specify the existence of several paths on the whole (\cite{FLS98,CGLV00, DT01}), nested regular expressions where multiple RPQs are organized into a tree structure (\cite{PAG10, BPR12}), extended CRPQs where regular or rational relations over paths are allowed (\cite{BHLW10, BFL12}). 

RPQs have also been extended in another way, called regular data path queries (RDPQ), to incorporate data value comparisons between two nodes in a path (\cite{LV12a}). RDPQs are interpreted over data graphs, which extend graph databases by assigning a data value to every node. A RDPQ looks for a pair of nodes connected by a path whose sequence of data values and labels is accepted by a nondeterministic register automata (NRA). NRA is an extension of finite state automata, where a fixed number of registers are used to store the data values. Similarly to 2RPQs and CRPQs, RDPQs with inverse (2RDPQ), conjunctive RDPQs (CRDPQ), or conjunctive 2RDPQs (C2RDPQ), can also be defined.

Evaluation and containment are two basic problems for database query languages. These two problems have been investigated extensively for RPQs and 2RPQs, CRPQs and C2RPQs (see \cite{Bar13} for a survey). For RDPQs, the evaluation problem is PSPACE-complete,  and NLOGSPACE-complete in data complexity. On the other hand, the containment problem of RDPQs is undecidable, as a result of the undecidability of the inclusion problem of NRAs  (\cite{LV12a}). Since the containment and equivalence problem are essential for the optimization of queries, this undecidability result of RDPQs seems to undermine the validity of RDPQs as a fundamental formalism of path queries that combines the labelling and data constraints.

Our goal in this paper is to propose an alternative extension of 2RPQs with data value comparisons, called rigid regular data path queries with inverse (2RRDPQ), which, we believe, achieves a good balance between the expressive power and the computational properties (decidability and complexity). 

2RRDPQs are based on an automaton model also proposed in this paper, called nondeterministic rigid register automata (NRRA), where the data value comparisons are ``rigid'' in the sense that if the data value in the current position $x$ is compared to  a data value in some other position $y$, then by only using the labels (but not data values), the position $y$ can be uniquely determined from $x$. With the rigidity constraint, we are able to show that NRRAs enjoy nice properties as finite state automata, that is, NRRAs can be determinized, they are closed under all Boolean operations, the two-way variant of NRRAs is expressively equivalent to (one-way) NRRAs, and they are expressively equivalent with a natural extension of regular expressions. In addition, while the expressive power of NRRAs and NRAs are incomparable, we demonstrate that the expressive power of NRRAs can be captured by an extension of NRAs with nondeterministic guessing.

To justify the expressibility of 2RRDPQs, we show that although the expressive power of 2RRDPQs and 2RDPQs are incomparable, every 2RDPQ can in fact be turned into a 2RRDPQ if a localized transformation is applied to graph databases. By ``localized transformation'', we mean that the transformation is obtained by adding for each node $v$ a new node $n_v$ which is only connected to $v$ and the global topology of the original graph is preserved (see Section \ref{sec-rrdpq}).

We then investigate the computational properties of 2RRDPQs. We show that 2RRDPQs can be evaluated over data graphs with the same (data and combined) complexity as RDPQs. In addition, we consider conjunctive 2RRDPQs (C2RRDPQ) and show that the containment problem of C2RRDPQs can be decided in 2EXPSPACE. From this, we deduce that the containment problem of 2RRDPQs can be decided in 2EXPSPACE as well. The 2EXPSPACE result is proved by a nontrivial extension of the proof for the EXPSPACE-completeness result of C2RPQs in \cite{CGLV00}, and is the most technical part of this paper.

\smallskip

\noindent {\it Related work}. The idea of rigidity is inspired by event clock automata from the verification community (\cite{AFH99}), where for every event $a$, a clock $x_a$ is used to record the time that has been elapsed from the last occurrence of $a$, and a clock $y_a$ is used to predict the time that will elapse until the next occurrence of $a$. Although in spirit similarly to event clock automata, NRRAs are defined to allow much more complicated data value comparisons. For instance, in NRRAs, the current data value can be compared to the data value in the position corresponding to the last occurrence of the word symbol $a$ before the next occurrence of the symbol $b$. This capability of data value comparisons is essential for the proof of the 2EXPSPACE result of the containment problem of C2RRDPQs (see Section 5). NRAs were introduced in \cite{KF94}. A restriction of NRAs, window memory automata, has been proposed in \cite{BLP10}, where only local data value comparisons are allowed. NRRAs strictly extend window memory automata, since non-local data value comparisons are allowed. Various query formalisms have been proposed for data graphs to combine the topology and data constraints, e.g. XPath with data comparisons \cite{KRV14}, TriAL for RDF \cite{LRV13}.  Although the containment problem of data path queries is in general undecidable, it has been examined in detail for various fragments with positive data value comparisons (\cite{KRV14}). 

\smallskip

\noindent {\it Organization of this paper}. Definitions are given in the next section. NRRAs and their variants are presented in Section 3. 2RRDPQs are investigated in Section 4. Section 5 deals with the C2RRDPQs.

\vspace*{-2mm}

\section{Definitions}
For a natural number $k$ such that $k > 0$, let $[k]$ denote $\{1,\dots,k\}$ and $[-k]$ denote $\{-k,\dots,-1\}$. 

Fix a finite alphabet $\Sigma$ and an infinite set of data values $\Dd$. Let $\Sigma^{\pm}=\Sigma \cup \{a^-\mid a \in \Sigma\}$. For $a \in \Sigma^{\pm}$, we use $a^{-}$ to denote the inverse of $a$. In particular, $(a^-)^-=a$ for $a \in \Sigma$.

A word $w$ over the alphabet $\Sigma$ is a finite sequence of elements from $\Sigma$. For a word $w$, $|w|$ is used to denote the length of $w$. 


A \emph{data path} $\alpha$ over $\Sigma$ is a sequence $d_0 a_1 d_1 \dots a_n d_n$, where $d_0,\dots,d_n \in \Dd$, and $a_1,\dots,a_n \in \Sigma$. The data path of the minimum length is a single data value $d$. Given two data paths $\alpha_1=d_0 a_1 d_1 \dots a_n d_n$ and $\alpha_2=d_n a_{n+1} d_{n+1}\dots a_m d_m$, the \emph{concatenation} of $\alpha_1$ and $\alpha_2$, denoted by $\alpha_1 \cdot \alpha_2$, is defined as the following data path, $d_0 a_1 d_1 \dots a_n d_n a_{n+1} d_{n+1}\dots a_m d_m$. Note that $\alpha_1 \cdot \alpha_2$ is defined only if the last data value of $\alpha_1$ is the same as the first data value of $\alpha_2$. The definition naturally extends to the concatenation of multiple data paths. 


A \emph{language} over $\Sigma$ is a set of words over $\Sigma$ and a \emph{data language} over $\Sigma$ is a set of data paths over $\Sigma$.

Let $\Sigma$ and $\Gamma$ be two finite alphabets. Then a \emph{letter projection} $prj$ from $\Sigma$ to $\Gamma$ is a \emph{surjective} function from $\Sigma$ to $\Gamma$. The letter projections of words, data paths, languages and data languages can be defined in a natural way. For a letter projection $prj: \Sigma \rightarrow \Gamma$ and $\gamma \in \Gamma$, we use $prj^{-1}(\gamma)$ to denote the set $\{a \in \Sigma \mid prj(a)=\gamma\}$. Note that $(prj^{-1}(\gamma))_{\gamma \in \Gamma}$ forms a partition of $\Sigma$. In addition, for $B \subseteq \Gamma$, let $prj^{-1}(B)=\bigcup \limits_{\gamma \in B} prj^{-1}(\gamma)$.

A \emph{graph database} $\gdb$ is an edge-labeled graph $(V,E)$, where $V$ is the set of nodes and $E \subseteq V \times \Sigma \times V$. For $e=(v,a,v') \in E$, let $\lambda(e)$ denote the label $a$. A \emph{semipath} $\pi$ in $\gdb$ is a sequence $v_0 a_1 v_1 \dots v_{n-1} a_n v_n$ such that for every $i: 1 \le i \le n$, either $(v_{i-1},a_i,v_i) \in E$ or $(v_i,a^{-}_i,v_{i-1}) \in E$. A \emph{path} $\pi$ in $\gdb$ is a semipath $v_0 a_1 v_1 \dots v_{n-1} a_n v_n$ such that for every $i: 1 \le i \le n$, $(v_{i-1},a_i,v_i) \in E$. Let $\lambda(\pi)$ denote the sequence of labels on a semipath $\pi$, that is, $a_1 \dots a_n$.  A semipath $\pi$ is \emph{simple} if no nodes are repeated on $\pi$.

A \emph{regular path query} (RPQ) over $\Sigma$ is a tuple $\xi=(x,L,y)$, where $L$ is a regular language over the alphabet $\Sigma$. The regular language $L$ can be given by a finite state automaton or a regular expression. Given a graph database $\gdb=(V,E)$, the evaluation result of $\xi$ over $\gdb$, denoted by $\xi(\gdb)$, consists of the set of pairs $(v,v')$ such that there is a path $\pi$ from $v$ to $v'$ such that $\lambda(\pi) \in L$. 

A \emph{regular path query with inverse} (2RPQ) over $\Sigma$ is a tuple $\xi=(x,L,y)$, where $L$ is a regular language over the alphabet $\Sigma^{\pm}$. The semantics of 2RPQs are defined similarly to RPQs, with paths replaced by semipaths.

The evaluation problem for a RPQ or 2RPQ is defined as follows: Given a RPQ or 2RPQ $\xi$, a graph database $\dg=(V,E,\eta)$, a node pair $(v,v')$ in $\dg$, decide whether $(v,v') \in \xi(\dg)$.

The containment problem of a RPQ or 2RPQ is defined as follows: Let $\xi_1,\xi_2$ be two RPQs or 2RPQs. Then $\xi_1$ is contained in $\xi_2$, denoted by $\xi_1 \subseteq \xi_2$, if for every graph database $\dg$, $\xi_1(\dg) \subseteq \xi_2(\dg)$.

A \emph{conjunctive regular path query} (CRPQ) $\xi$ over $\Sigma$ is an expression of the form
$Ans(\bar{z}) \leftarrow \bigwedge \limits_{1 \le i \le l} (y_{2i-1}, L_i, y_{2i})$, where for every $i$, $(y_{2i-1},L_i,y_{2i})$ is a RPQ over $\Sigma$, and $\bar{z}$ is a tuple of variables from $\{y_1,\dots,y_{2l}\}$ ($\bar{z}$ are called the \emph{distinguished variables} of $\xi$). Note that in the above definition, $y_i$ and $y_j$ ($i \neq j$) may be the same variable. 

Given a graph database $\dg=(V,E)$, a CRPQ $\xi: = Ans(\bar{z}) \leftarrow \bigwedge \limits_{1 \le i \le l} (y_{2i-1}, L_i, y_{2i})$, and $\nu: \{y_1,\dots,y_{2l}\} \rightarrow V$, we say $(\dg,\nu) \models \xi$, if $(\nu(y_{2i-1}),\nu(y_{2i}))$ belongs to the evaluation result of $(y_{2i-1},L_i,y_{2i})$ over $\dg$ for every $i: 1 \le i \le l$. The evaluation result of $\xi$ over $\dg$, denoted by $\xi(\dg)$, is the set of all tuples $\nu(\bar{z})$ for $\nu: \{y_1,\dots,y_{2l}\} \rightarrow V$ such that $(\dg,\nu) \models \xi$. Similarly, C2RPQs can be defined, with RPQs replaced by 2RPQs.



The evaluation and containment problem of CRPQs or C2RPQs can be defined similarly to RPQs.

A \emph{data graph} $\dg$ is a tuple $(V,E,\eta)$, where $(V,E)$ is a graph database and $\eta: V \rightarrow \Dd$ assigns each node a data value. For a semipath $\pi=v_0 a_1 v_1 \dots v_{n-1} a_n v_n$ in $(V,E)$, the \emph{data path} corresponding to $\pi$, denoted by $\eta(\pi)$, is $\eta(v_0) a_1 \eta(v_1)\dots \eta(v_{n-1}) a_n \eta(v_n)$. 


Let $k$ be a natural number. A \emph{$k$-register data constraint} is defined by the following rules: 
\[c:=r_i \sim r_j \mid r_i \nsim r_j \mid c \vee c \mid c \wedge c, \]
where $0 \le i, j \le k$, $i \neq j$, and $r_0$ is a special register reserved for the current data value.
Let $\Cc_k$ denote the set of data constraints.


Let $c$ be a data constraint, $\theta \in (\Dd \cup \{\bot\})^{[k]}$ be the current state of registers (where $\theta(i)=\bot$ denotes the fact that no data value is stored into the register $r_i$), and $d$ be a data value, then the semantics of $c$ is defined over $(\theta,d)$ as follows:
If $c=r_i \sim r_j$, then $(\theta,d) \models c$ iff $\theta[0 \leftarrow d](i) \neq \bot$, $\theta[0 \leftarrow d](j)\neq \bot$, and $\theta[0 \leftarrow d](i)=\theta[0 \leftarrow d](j)$, where $\theta[0 \leftarrow d]$ is the function extending $\theta$ by assigning $d$ to $0$. The semantics of $c=r_i \nsim r_j$ can be defined similarly. In addition, the semantics of $c=c_1 \vee c_2$ and $c=c_1 \wedge c_2$ are defined in a natural way.


Let $k$ be a natural number. A \emph{nondeterministic $k$-register data path automaton} (NRA, \cite{LV12a}) $\Aa$ over $\Sigma^{\pm}$ is a tuple $(Q,k,\delta,I,F)$, where $Q=Q_w \cup Q_d$ such that $Q_w$ and $Q_d$ are two finite disjoint sets of word states and data states,  $k$ is the number of registers, $I \subseteq Q_d$ is the set of initial states, $F \subseteq Q_w$ is the set of final states, $\delta =\delta_w \cup \delta_d$ such that $\delta_w \subseteq Q_w \times \Sigma^{\pm} \times Q_d$ is the word transition relation and $\delta_d \subseteq Q_d \times \Cc_k \times Q_w \times 2^{[k]}$ is the data transition relation.

The intuition of the definition of NRAs is that since data paths alternate between data values and word symbols, when $\Aa$ is in a data (resp. word) state, it is ready to read a data value (resp. a word symbol). Since data paths begin and end with data values, an initial state should be a state before reading a data value, so $I$ is defined as a subset of $Q_d$, dually, a final state should be a state after reading a data value, so $F$ is defined as a subset of $Q_w$.

Given a data path $\alpha=d_0 a_1 d_1 \dots a_n d_n$ and a NRA $\Aa=(Q,k,\delta,I,F)$, a \emph{configuration} of $\Aa$ on $\alpha$ is a tuple $(q,j,\theta)$, where $q \in Q$, $j$ is the current position (where $j=0$ means the first position) of the symbol that $\Aa$ reads, and $\theta \in (\Dd \cup \{\bot\})^{[k]}$ is the current state of the registers. An \emph{initial configuration} of $\Aa$ over $\alpha$ is $(q,0,\theta_\bot)$, where $q \in I$, and $\theta_\bot(i)=\bot$ for every $i \in [k]$. Let $(q,j,\theta),(q',j+1,\theta')$ be two configurations (where $0 \le j \le 2n$). Then $(q',j+1,\theta')$ is said to be a \emph{successor} of $(q,j,\theta)$, denoted by $(q,j,\theta) \vdash_\alpha (q',j+1,\theta')$,  if one of the following conditions holds.
\begin{itemize}
\item If the $j$-th symbol of $\alpha$ is a word symbol $a$, then $(q,a,q')\in \delta_w$, $\theta'=\theta$.
\item If the $j$-th symbol of $\alpha$ is a data value $d$, then there are $c \in \Cc_k, X \subseteq [k]$ such that $(q,c,q',X)\in \delta_d$, $(\theta,d) \models c$, and $\theta'$ is obtained from $\theta$ by assigning $d$ to every $i \in X$.
\end{itemize} 

A data path $\alpha=d_0 a_1 d_1 \dots a_n d_n$ is \emph{accepted} by a NRA $\Aa$ if there are $q \in I, q' \in F$ and a data assignment $\theta$ such that $(q,0,\theta_\bot) \vdash_\alpha^\ast (q',2n+1,\theta)$, where $\vdash_\alpha^\ast$ is the reflexive and transitive closure of $\vdash_\alpha$. The set of data paths accepted by $\Aa$ is denoted by $\Ll(\Aa)$. In addition, for every $\theta,\theta' \in (\Dd \cup \{\bot\})^{[k]}$, we use $\Ll(\Aa,\theta,\theta')$ to denote the set of data paths $\alpha=d_0 a_1 d_1 \dots a_n d_n$ such that there are $q \in I, q' \in F$ satisfying $(q,0,\theta) \vdash_{\alpha}^\ast (q',2n+1,\theta')$.

Let $k$ be a natural number. \emph{Regular expressions with $k$-memory} (REM, \cite{LV12a}) over $\Sigma^{\pm}$ are defined by the following rules:
\[e:=\varepsilon\mid \emptyset \mid  a \mid  e \cdot e \mid e \cup e \mid e^+ \mid \ \downarrow_X e \mid e [c],\]
where $a \in \Sigma^\pm$, $c\in \Cc_k$, and $X \subseteq [k]$.


The semantics of REMs is defined by a relation $\theta \vdash_{e,\alpha} \theta'$, where $e$ is a REM, $\alpha$ is a data path, $\theta,\theta' \in (\Dd \cup \{\bot\})^{[k]}$. In the following, due to space constraints, we only present the semantics for the last two rules above, that is, $ e=\downarrow_X e_1$ and $e=e_1[c]$,  the semantics of the other rules are obvious and can be found in \cite {LV12a}.
\begin{itemize}
%
%
%
%
%
\item If $e=\downarrow_X e_1$, then $\theta \vdash_{e,\alpha} \theta'$ if $\theta_{X=d} \vdash_{e_1,\alpha} \theta'$, where $d$ is the first data value of $\alpha$, and $\theta_{X=d}$ is obtained from $\theta$ by assigning $d$ to all the registers in $X$.
\item If $e=e_1 [c]$, then $\theta \vdash_{e,\alpha} \theta'$ if $\theta \vdash_{e_1,\alpha} \theta'$ and $(\theta',d)\models c$, where $d$ is the last data value of $\alpha$.
\end{itemize}

A data path $\alpha$ is accepted by a REM $e$ if there exists $\theta \in (\Dd \cup \{\bot\})^{[k]}$ such that $\theta_\bot \vdash_{e,\alpha}\theta$. The set of data paths accepted by a REM $e$ is denoted by $\Ll(e)$. For every $\theta,\theta' \in (\Dd \cup \{\bot\})^{[k]}$, we use $\Ll(e,\theta,\theta')$ to denote the set of data paths $\alpha$ such that $\theta \vdash_{e,\alpha}\theta'$.


\vspace*{-2mm}
\begin{theorem}[\cite{DL09,LV12a,LV12b}]
The following facts hold for NRAs and REMs.
\begin{itemize}
\item NRAs and REMs are expressively equivalent.
\item The nonemptiness problem of NRAs and REMs is PSPACE-complete.
\item The universality and equivalence problem of NRAs and REMs are undecidable.
\end{itemize}
\end{theorem}
\vspace*{-2mm}
A \emph{regular path query with data}(RDPQ) $\xi$ over $\Sigma$ is a tuple $(x, L,y)$, where $L$ is a language of data paths defined by a NRA or a REM over the alphabet $\Sigma$. Given a data graph $\dg=(V,E,\eta)$, the evaluation result  of $\xi$ over $\dg$, denoted by $\xi(\dg)$, is the set of node pairs $(v,v')$ in $\dg$ such that there is a path $\pi$ from $v$ to $v'$ such that $\eta(\pi)$, the data path corresponding to $\pi$, belongs to $L$.

A \emph{regular path query with inverse and data} (2RDPQ) $\xi$ over $\Sigma$ is a tuple $(x, L,y)$, where $L$ is a language of data paths defined by a NRA or a REM over $\Sigma^\pm$. The semantics of 2RDPQ $\xi$ over a data graph $\dg=(V,E,\eta)$ is defined similarly to that of RDPQ, with paths replaced by semipaths.

Similarly to 2RPQs, regular path queries with inverse and data (2RDPQ) can be defined. Moreover, CRDPQs and C2RDPQs can be defined in the same way as CRPQs and C2RPQs. The evaluation and containment problem of RDPQs, 2RDPQs, CRDPQs, C2RDPQs can also be defined similarly.


\vspace*{-2mm}

\begin{theorem}[\cite{LV12a}]
The following results hold for RDPQs, 2RDPQs, CRDPQs and C2RDPQs.
\begin{itemize}
\item The evaluation problem of RDPQs and 2RDPQs is PSPACE-complete, and NLOGSPACE-complete in data complexity.
\item The evaluation problem of CRDPQs and C2RDPQs is PSPACE-complete, and NLOGSPACE-complete in data complexity.
\item The containment problem of RDPQs, 2RDPQs, CRDPQs and C2RDPQs is undecidable. 
\end{itemize}
\end{theorem}


\vspace*{-4mm}

\section{Rigid register automata and its relatives}

In this section, we first define nondeterministic rigid register automata (NRRA). Then we show the robustness of this model by proving that NRRA can be determinized and their two-way as well as alternating variants are expressively equivalent to NRRA. We also show that there is a natural extension of regular expressions equivalent to NRRA.

\vspace*{-1mm}

\subsection{Rigid data constraints}

A \emph{position term} $t$ over the alphabet $\Sigma^\pm$ is defined by the following rules,
\[t:=\cur \mid \suc(t) \mid \pred(t) \mid \suc_A(t) \mid \pred_A(t),\]
where $A$ is a nonempty subset of $\Sigma^{\pm}$. Intuitively,  the constant ``$\cur$'' denotes the position of the current data value, 
``$\suc$'' and ``$\pred$'' denote the position of the next and the previous data value, `$\suc_A$'' denotes the position of the data value \emph{immediately after the next occurrence of a word symbol from $A$}, dually, ``$\pred_A$'' denotes the position of the data value \emph{immediately before the previous occurrence of a word symbol from $A$}.

Let $\Tt_p[\Sigma^\pm]$ denote the set of position terms over $\Sigma^\pm$. 

For briefness, position terms of the form $\suc_{\{a\}}(t)$ or $\pred_{\{a\}}(t)$ (where $a \in \Sigma^\pm$) are written as $\suc_a(t)$ or $\pred_a(t)$. In addition, we use $\suc^i$ to denote the repetitions of $\suc$ for $i$ times. Similarly, we use the abbreviations $\pred^i$, $\suc^i_A$, and $\pred^i_A$.

The set of subterms of $t \in \Tt_p[\Sigma^\pm]$, denoted by $sub(t)$, are defined in a natural way, e.g. $sub(\suc_A(t_1))=\{\suc_A(t_1)\} \cup sub(t_1)$. We use $t' \preceq t$ to denote the fact that $t' \in sub(t)$, and $t' \prec t$ to denote the fact that $t' \preceq t$ and $t \neq t'$. Suppose $t,t',t_1 \in \Tt_p[\Sigma^\pm]$ and $t' \preceq t$, let $t[t' \backslash t_1]$ denote the position term obtained from $t$ by replacing $t'$ with $t_1$. 

The semantics of position terms are defined as follows: Give a data path $\alpha=d_0 a_1 d_1 \dots a_n d_n$ and a position $2i$ (where $0 \le i \le n$, $2i$ is the position for the data value $d_i$, and the first position is indexed by $0$), the position represented by $t$ over $\alpha$ and $2i$,  denoted by $t_{\alpha}[2i]$, is defined as follows.
\begin{itemize}
\item  $\cur_{\alpha}[2i]=2i$.
\item If $i < n$, then $(\suc(\cur))_{\alpha}[2i]=2(i+1)$. Otherwise, $(\suc(\cur))_{\alpha}[2i] =\bot$.
\item If $i>0$, then $(\pred(\cur))_{\alpha}[2i]=2(i-1)$. Otherwise, $(\pred(\cur))_{\alpha}[2i]=\bot$.
\item If ${(t_1)}_{\alpha}[2i] \neq \bot$, then
\[(\suc(t_1))_{\alpha}[2i]=(\suc(\cur))_{\alpha}[{(t_1)}_{\alpha}[2i]].\]
Otherwise, $(\suc(t_1))_{\alpha}[2i] = \bot$.
\item If $(t_1)_{\alpha}[2i] \neq \bot$, then 
\[(\pred(t_1))_{\alpha}[2i]=(\pred(\cur))_{\alpha}[{(t_1)}_{\alpha}[2i]].\]
Otherwise, $\pred(t_1))_{\alpha}[2i] = \bot$,
\item If there exists $j: i < j \le n$ such that $a_{j} \in A$ and $j$ is the minimum number satisfying this condition, that is, for every $j': i < j' < j$, we have $a_{j'} \nin A$,  then $(\suc_A(\cur))_{\alpha}[2i]=2j$. Otherwise, $(\suc_A(\cur))_{\alpha}[2i]=\bot$. 
\item If there exists $j: j \le i$ such that $a_{j} \in A$ and $j$ is the maximum number satisfying this condition, that is,  for every $j': j < j' \le i$, we have $a_{j'} \nin A$, then $(\pred_A(\cur))_{\alpha}[2i]=2(j-1)$. Otherwise, $(\pred_A(\cur))_{\alpha}[2i]=\bot$.
\item If ${(t_1)}_{\alpha}[2i] \neq \bot$, then 
\[(\suc_A(t_1))_{\alpha}[2i]=(\suc_A(\cur))_{\alpha}[{(t_1)}_{\alpha}[2i]].\]
Otherwise, $(\suc_A(t_1))_{\alpha}[2i]=\bot$.
\item If ${(t_1)}_{\alpha}[2i] \neq \bot$, then 
\[(\pred_A(t_1))_{\alpha}[2i]=(\pred_A(\cur))_{\alpha}[{(t_1)}_{\alpha}[2i]].\]
Otherwise, $(\pred_A(t_1))_{\alpha}[2i]=\bot$.
\end{itemize}


\vspace*{-4mm}
\begin{example}
Suppose 
\[\alpha=\begin{array}{c c c c c c c c c c c }d_0 & a & d_1 & b & d_2 & a & d_3 & a & d_4 & b & d_5 \\
0 & 1 & 2 & 3 & 4 & 5 & 6 & 7 & 8 & 9 & 10\end{array},\] 
where the second arrow is the sequence of positions. Let $t_1=\suc_{a}(\suc(\cur))$ and $t_2=\pred_{b}(\pred(\cur))$. Let us consider $(t_1)_{\alpha}[0]$ and $(t_2)_{\alpha}[10]$. At first, $(t_1)_{\alpha}[0]=(\suc_{a}(\cur))_{\alpha}[(\suc(\cur))_\alpha[0]]$. Since 
$(\suc(\cur))_{\alpha}[0]=2$, and the first occurrence of $a$ after the position $2$ is in the position $5$, we get $(t_1)_\alpha[0]=6$. On the other hand, $(t_2)_{\alpha}[10]=(\pred_{b}(\cur))_\alpha[(\pred(\cur))_\alpha[10]]$. Because $(\pred(\cur))_\alpha[10]=8$ and the last occurrence of $b$ before the position $8$ is in the position $3$, we get $(t_2)_\alpha[10]=2$.
\end{example}


\vspace*{-1mm}
A \emph{rigid data constraint} $c$ over $\Sigma^\pm$ is defined by the following rules, 
\[c:= t_1 \sim t_2 \mid t_1 \nsim t_2 \mid c  \vee c \mid c \wedge c, \mbox{ where } t_1,t_2 \in \Tt_p[\Sigma^\pm].\]

We use $\Cc_{rgd}[\Sigma^\pm]$ to denote the set of rigid data constraints over $\Sigma^\pm$.

The semantics of rigid data constraints can be defined inductively. In the following, we will define the semantics for the case $c=t_1 \sim t_2$. The semantics of $c=t_1 \nsim t_2$ can be defined similarly. Moreover, the semantics of $c=c_1 \vee c_2$ and $c=c_1 \wedge c_2$ can be defined in a standard way. Let $c \in \Cc_{rgd}[\Sigma^\pm]$, $\alpha=d_0 a_1 d_1 \dots a_n d_n$, and $i: 0 \le i \le n$, then $(\alpha,2i)$ is said to satisfy $c=t_1 \sim t_2$, denoted by $(\alpha,2i) \models c$, if ${(t_1)}_{\alpha}[2i] \neq \bot$, ${(t_2)}_{\alpha}[2i] \neq \bot$, and $d_{{(t_1)}_{\alpha}[2i]} = d_{{(t_2)}_{\alpha}[2i]}$.

Given a rigid data constraint $c$, we use $\bar{c}$ to denote the negation of $c$. More specifically, $\bar{c}$ is obtained from $c$ by swapping $\sim$ for $\nsim$, and $\vee$ for $\wedge$. For instance, if $c=\cur \sim \suc_a(\cur) \vee \cur \nsim \pred(\cur)$, then $\overline{c}=\cur \nsim \suc_a(\cur) \wedge \cur \sim \pred(\cur)$.

\vspace*{-2mm}

\begin{proposition}\label{prop-sat-rdc}
The satisfiability problem of rigid data constraints is NP-complete.
\end{proposition}

\vspace*{-4mm}

\subsection{Nondeterministic and deterministic rigid register automata}

A \emph{nondeterministic rigid register automaton (NRRA)} $\Aa$ over the alphabet $\Sigma^\pm$ is a tuple $(Q,\delta,I,F)$, where $Q,I,F$ are as those in NRA, $\delta = \delta_w \cup \delta_d$ such that $\delta_w \subseteq Q_w \times \Sigma^\pm \times Q_d$ and $\delta_d \subseteq Q_d \times \Cc_{rgd}[\Sigma^\pm] \times Q_w$.

A \emph{run} of $\Aa$ over a data path $\alpha=d_0 a_1 d_1 \dots a_n d_n$ is a state sequence $q_0 c_0 q_1 a_1 q_2 \dots q_{2n-1} a_n q_{2n} c_n q_{2n+1}$ such that $q_0 \in I$,  for every $i: 0 \le i \le n$, $(q_{2i},  c_i, q_{2i+1}) \in \delta_d$ and $(\alpha,2i) \models c_i$, and for every $i: 1 \le i \le n$, $(q_{2i-1},a_i,q_{2i}) \in \delta_w$. A run $\rho=q_0 c_0 q_1 a_1 q_2 \dots q_{2n-1} a_n q_{2n} c_n q_{2n+1}$ is \emph{accepting} if $q_{2n+1} \in F$.

A \emph{deterministic} rigid register automaton (DRRA) over $\Sigma^\pm$ is a NRRA $\Aa=(Q,\delta,I,F)$ such that $I$ is a singleton, and $\delta$ satisfies that for every $q \in Q_w, a \in \Sigma^\pm$, there is at most one $q' \in Q_d$ such that $(q,a,q')\in \delta_w$, and for every $(q,c_1,q_1), (q,c_2,q_2) \in \delta_d$, if $q_1 \neq q_2$, then $c_1 \wedge c_2$ is unsatisfiable. 

Let $\Aa$ be a NRRA. Then $\Tt_\Aa$ is used to denote the minimal set of position terms satisfying that for every $t_1 \sim t_2$ or $t_1 \nsim t_2$ occurring in $\Aa$, we have $t_1,t_2 \in \Tt_\Aa$; moreover, if $t \in \Tt_\Aa$ and $t' \preceq t$, then $t', t[t' \backslash \cur] \in \Tt_\Aa$. In addition, $\Cc_\Aa$ is used to denote the set of rigid data constraints occurring in $\Aa$.

\begin{figure}[htbp]
\begin{center}
\includegraphics[width=0.45\textwidth]{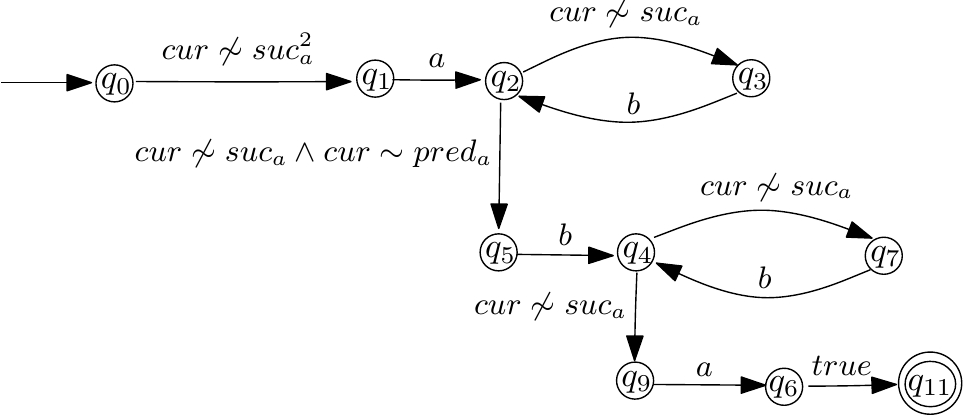}
\end{center}
\caption{An example for NRRA}
\label{fig-rra-example}
\end{figure}

\vspace*{-2mm}
\begin{example}
Let $\Sigma=\{a,b\}$. Let $L$ denote the language of data paths satisfying that the sequence of word symbols on the data path belongs to $ab^\ast a$, the first data value occurs in some other position, and the last data value does not occur elsewhere.
Then $L$ is defined by the NRRA $\Aa$ illustrated in Figure \ref{fig-rra-example}, where $\suc_a$ is an abbreviation of $\suc_a(\cur)$, $Q_d=\{q_0,q_2,q_4,q_6\}$ and $Q_w=\{q_1,q_3,q_5,q_7,q_9,q_{11}\}$. 
\end{example}

\vspace*{-1mm}

Since NRRAs are able to compare the current data value with the data values in the future, NRAs and NRRAs are expressively incomparable.

\vspace*{-2mm}
\begin{proposition}\label{prop-expr-nra-nrra}
NRA and NRRA are expressively incomparable.
\end{proposition}


\vspace*{-2mm}

Let $\Aa=(Q,\delta,I,F)$ be a NRRA over the alphabet $\Sigma^\pm$ and $prj$ a letter projection from $\Sigma^\pm$ to $\Gamma$. Then the \emph{letter projection} of $\Aa$, denoted by $prj(\Aa)$, is obtained from $\Aa$ by replacing each transition $(q,a,q') \in \delta_w$ with $(q,prj(a),q')$, and each $\suc_A$ (resp. $\pred_A$) occurring in $\delta_d$, where $A \subseteq \Sigma^\pm$, with $\suc_{prj(A)}$ (resp. $\pred_{prj(A)}$). Note that $prj(\Aa)$ may not define $prj(\Ll(\Aa))$, as witnessed by the following result.

\vspace*{-2mm}
 
\begin{proposition}\label{prop-let-proj}
The class of languages definable by NRRAs are not closed under letter projections.
\end{proposition}
\vspace*{-1mm}

In the following, we will introduce a constraint for a pair $(\Aa,prj)$, where $\Aa$ is a NRRA and $prj$ is a letter projection, so that $prj(\Aa)$ does define $prj(\Ll(\Aa))$. 

Let $\Aa=(Q,\delta,I,F)$ be a NRRA over $\Sigma^\pm$, $prj$ be a letter projection from $\Sigma^\pm$ to $\Gamma$. Then $\Aa$ is said to be \emph{position-invariant} under $prj$ if for every $\suc_A$ (resp. $\pred_A$) occurring in $\delta_d$, where $A \subseteq \Sigma^\pm$, there is $B \subseteq \Gamma$ such that $A=prj^{-1}(B)$. It is easy to observe that the position-invariance guarantees that for every $t \in \Tt_\Aa$, every data path $\alpha$ and every position $2i$ of $\alpha$, it holds $t_\alpha[2i]=(prj(t))_{prj(\alpha)}[2i]$, where $prj(t)$ is obtained from $t$ by replacing each occurrence of $\suc_A$ (resp. $\pred_A$) in $t$ with $\suc_{prj(A)}$ (resp. $\pred_{prj(A)}$). From this, we deduce that if $\Aa$ is position invariant under $prj$, then $prj$ does not affect the interpretations of the rigid data constraints in $\Aa$. So we have the following result.

\vspace*{-2mm}
\begin{proposition}\label{prop-nrra-pos-inv}
Suppose $\Aa$ is a NRRA over $\Sigma^\pm$ and $prj$ is a letter projection from $\Sigma^\pm$ to $\Gamma$. If $\Aa$ is position-invariant under $prj$, then $\Ll(prj(\Aa))=prj(\Ll(\Aa))$.
\end{proposition}
\vspace*{-1mm}

For a NRA, in every position, only a bounded number of data values occurring before this position are stored into the registers  for the future references. On the other hand, in the first sight, in a NRRA, it is only required that a bounded number of positions are referenced to by a data transition in a \emph{single} position, but it is not required that only a bounded number of positions are referenced to by \emph{all} the data transitions after a position. In the following, we show that this is indeed the case. By utilizing this property, we then show that NRRAs can be simulated by an extension of NRAs with nondeterministic guessing\footnote{The idea of nondeterministic guessing,  called nondeterministic reassignment, was introduced in \cite{KZ08}.}.

Let $\Aa=(Q,\delta,I,F)$ be a NRRA over the alphabet $\Sigma^\pm$, $\alpha=d_0 a_1 d_1 \dots a_n d_n$ be a data path over $\Sigma^\pm$, $\rho=q_0 c_0 q_1 a_1 q_2 \dots q_{2n-1} a_n q_{2n} c_n q_{2n+1}$ be a run of $\Aa$ over $\alpha$, and $i: 0 \le i \le n$. Define the set of \emph{future positions} of the position $2i$ of $\alpha$ with respect to $\rho$, denoted by $Pos^f_{\rho}[\alpha,2i]$, as 
\[\{t_{\alpha}[2j] \mid j \le i, t \mbox{ occurs in } c_j, t_\alpha[2j] \neq \bot, t_\alpha[2j]> 2i\}.\]
Similarly, define the set of \emph{past positions} of the position $2i$ of $\alpha$ with respect to $\rho$, denoted by $Pos^p_{\rho}[\alpha,2i]$, as 
\[\{t_{\alpha}[2j] \mid i \le j, t \mbox{ occurs in } c_j, t_\alpha[2j] \neq \bot, t_\alpha[2j]< 2i\}.\]

\vspace*{-3mm}
\begin{lemma}\label{lem-nrra-pos-bnd}
Let $\Aa=(Q,\delta,I,F)$ be a NRRA over the alphabet $\Sigma^\pm$ and $\alpha=d_0 a_1 d_1 \dots a_n d_n$ be a data path. Then for every run $\rho$ of $\Aa$ over $\alpha$ and every $i: 0 \le i \le n$, $Pos^f_{\rho}[\alpha,2i] \cup Pos^p_{\rho}[\alpha,2i] \subseteq \{t_\alpha[2i] \mid t \in \Tt_\Aa\}$. 
\end{lemma}
\vspace*{-2mm}

Intuitively, Lemma \ref{lem-nrra-pos-bnd} says that for every run $\rho$ over a data path $\alpha$ and every position $2i$ of $\alpha$, only a bounded number of positions before (resp. after) the position $2i$ are referred to by $\rho$ after (resp. before) reaching the position $2i$.

A \emph{nondeterministic register data path automaton with guessing} (NRAG) $\Aa$ over $\Sigma^\pm$ is a tuple $(Q,k,\delta,I,F)$, where  $Q, k, I ,F$ are as those in the definition of NRA,
$\delta \subseteq \delta_w \cup \delta_d$ such that $\delta_w \subseteq Q_w \times \Sigma^\pm \times Q_d$ and $\delta_d \subseteq Q_d \times \Cc_{k} \times Q_w \times 2^{[k]} \times 2^{[k]} \times \Cc_{2k}$ satisfies that for every $(q,c,q',X,Y,c') \in \delta_d$, it holds that $X \cap Y = \emptyset$, and $c'$ \emph{does not} contain $r_{k+i}$ with $r_i \nin Y$.

The intuition of a transition $(q,c,q',X,Y,c') \in \delta$ is that if the current state is $q$, the data values stored in the registers together with the current data value $d$ satisfies $c$, then the state is changed to $q'$, $d$ is stored into every register in $X$. Meanwhile, for each register in $Y$, a data value is nondeterministically guessed. In addition, the guessed data values should satisfy the constraint $c'$.

The semantics of NRAGs are defined similarly as those of NRAs, that is, a successor relation of configurations $(q,j,\theta) \vdash_\alpha (q',j+1,\theta)$ can be defined, with the following adjustment for data transitions.
\begin{quote}
If the $j$-th symbol is a data value $d$, then there exist $c,c'$ such that $(q,c,q',X,Y,c') \in \delta_d$, and $\theta'$ is obtained from $\theta$ as follows,
\vspace*{-1mm}
\begin{itemize}
\item for each $i \in X$, $d$ is assigned to $i$ (thus $\theta'(i)=d$), 
\item for each $i \in Y$, a data value $d'_i$ is guessed (thus $\theta'(i)=d'_i$), so that the guessed data values satisfy the following condtion: The function $\theta_g$ extending $\theta$ by assigning $d'_i$ to $k+i$ for each $i \in Y$ satisfies that $(\theta_g,d) \models c'$,
\item for each $i \nin X \cup Y$, $\theta'(i)=\theta(i)$. 
\end{itemize}
\end{quote}
\vspace*{-2mm}

Note that data values are not allowed to be copied explicitly among  the registers in NRAG. But this can be achieved by guessing. For instance, if we want to \emph{copy} a data value from $r_i$ to $r_j$, then we can guess a data value for $r_j$ and add the constraint $r_i \sim r_{k+j}$ for the guessing. Later on, when we mention copying a data value from a register to the other, we always mean the implicit copying by guessing.

Since the nonemptiness of NRAGs can be solved similarly to that of NRAs, we have the following result.
\vspace*{-1mm}
\begin{proposition}\label{prop-nonempt-nrag}
The nonemptiness problem of NRAG is PSPACE-complete.
\end{proposition}
In the following, we will show that the expressive power of NRRAs can be captured by NRAGs.
\vspace*{-1mm}
\begin{theorem}\label{thm-nrra-nrag}
From a NRRA $\Aa=(Q,\delta,I,F)$, an equivalent NRAG $\Bb=(Q',k,\delta',I',F')$ can be constructed such that $|Q'|$ is polynomial over $|Q|$ and exponential over $|\Tt_\Aa|$ and $k$ is polynomial over $|\Tt_\Aa|$.
\end{theorem}


We will present a proof sketch for Theorem \ref{thm-nrra-nrag} and illustrate the main ideas. These ideas are also used for the proof of Theorem \ref{thm-rrdpq-auto} in Section 5.

\begin{proof}
Let $\Aa=(Q,\delta,I,F)$ be a NRRA. In the following, we will construct a NRAG $\Bb$ to simulate $\Aa$.

We first give an intuitive description of the construction. Let $\rho$ be a run of $\Aa$ over a data path $\alpha$. Then in the position $2i$, $\Bb$ simulates $\rho$ as follows: $\Bb$ records the data values in the positions belonging to $Pos^p_\rho[\alpha,2i]$, guesses the data values in the positions belonging to $Pos^f_\rho[\alpha,2i]$, and records the order for the positions in $Pos^p_\rho[\alpha,2i]$ and $Pos^f_\rho[\alpha,2i]$.


We introduce some additional notations.

Let $\alpha=d_0 a_1 d_1 \dots a_n d_n$ be a data path and $i: 0 \le  i \le n$.  The \emph{profile} of the position $2i$ in $\alpha$, denoted by $prof_{\alpha}(2i)$, is defined as a triple $(S,\chi, \sim)$, where
\vspace*{-1mm}
\begin{itemize}
\item $S=\{t \in \Tt_\Aa \mid t_{\alpha}[2i] \neq \bot\}$,
\item $\chi$ is a sequence
\[\begin{array}{l}(b_{-m_1}, T_{-m_1}, b'_{-m_1},s_{-m_1}) \dots (b_{-1}, T_{-1}, b'_{-1},s_{-1}) \\
(b_0, T_0,b'_0, s_{0}) (b_1, T_1,b'_1, s_1) \dots(b_{m_2}, T_{m_2},b'_{m_2},s_{m_2})\end{array}\]
where
\begin{itemize}
\item for every $j: -m_1 \le j \le m_2$, $T_j \subseteq S$ and $T_j \neq \emptyset$,
\item the collection $T_{-m_1},\dots,T_0, \dots,T_{m_2}$ forms a partition of $S$, and $\cur \in T_0$,
\item for every $t,t' \in S$, if $t \in T_{j_1}$ and $t' \in T_{j_2}$, then $j_1 \le j_2$ iff $t_{\alpha}[2i] \le t'_{\alpha}[2i]$ (in particular, $j_1 = j_2$ iff $t_{\alpha}[2i] = t'_{\alpha}[2i]$),
\item for every $j: -m_1 < j \le m_2$, $b_j=a_{(t_{\alpha}[2i])/2}$ for some $t \in T_j$, and $b_{-m_1}=a_{(t_{\alpha}[2i])/2}$ if $t_\alpha [2i]> 0$ for some $t \in T_{-m_1}$, otherwise, $b_{-m_1}=\bot$,
\item for every $j: -m_1 \le j < m_2$, $b'_j=a_{(t_{\alpha}[2i])/2+1}$ for some $t \in T_j$, and $b'_{m_2}=a_{(t_{\alpha}[2i])/2+1}$ if $t_\alpha [2i]< 2n$ for some $t \in T_{m_2}$, otherwise, $b'_{m_2}=\bot$,
\item $s_{m_2}=\bot$, and for every $j: -m_1 \le j < m_2$, if $t'_{\alpha}(2i)=t_{\alpha}(2i)+1$ for some $t \in T_j$ and $t' \in T_{j+1}$, then $s_j = 1$, otherwise, $s_j=0$.
\end{itemize}
\item $\sim$ is an equivalence relation over $S$ defined as follows: Let $t,t'\in S$, then $t \sim t'$ iff $d_{t_{\alpha}[2i]}=d_{t'_{\alpha}[2i]}$.
\end{itemize}
\vspace*{-1mm}

Let $\Sigma_{prof}$ denote the set of all triples $(S,\chi,\sim)$ such that $S \subseteq \Tt_\Aa$,
\begin{itemize}
\item $\chi$ is a sequence
\[\begin{array}{l}(b_{-m_1}, T_{-m_1}, b'_{-m_1}, s_{-m_1}) \dots (b_{-1}, T_{-1}, b'_{-1},s_{-1}) \\
(b_0, T_0,b'_0,s_{0}) (b_1, T_1,b'_1,s_1) \dots(b_{m_2}, T_{m_2},b'_{m_2},s_{m_2})\end{array}\]
 such that 
 \begin{itemize}
\item for every $j: -m_1 \le j \le m_2$, $T_j \neq \emptyset$, 
\item $\cur \in T_0$, and $T_{-m_1},\dots,T_{m_2}$ is a partition of $S$,
\item $b_{-m_1} \in \Sigma^\pm \cup \{\bot\}$, and for every $j: -m_1 < j \le m_2$, $b_j \in \Sigma^\pm$, 
\item $b'_{m_2} \in \Sigma^\pm \cup \{\bot\}$, and for every $j: -m_1\le j < m_2$, $ b'_j \in \Sigma^\pm$, 
\item $s_{m_2}=\bot$, and for every $j: -m_1 \le j < m_2$, $s_j \in \{0,1\}$,
\end{itemize}
\item $\sim$ is an equivalence relation over $S$ such that for every $t,t' \in \Tt_\Aa$, if $t,t' \in T_j$ for some $j$, then $t \sim t'$.
\end{itemize}
\vspace*{-1mm}
Note that for $(S,\chi, \sim) \in \Sigma_{prof}$, there may be no data paths  $\alpha$ and a position in $\alpha$ such that the profile of the position in $\alpha$ is $(S,\chi,\sim)$. Nevertheless, we are able to define a consistency condition on the elements from $\Sigma_{prof}$ so that a consistent element from $\Sigma_{prof}$ indeed corresponds to the profile of a position in some data path. Moreover, for two consistent elements from $\Sigma_{prof}$, say $(S_1,\chi_1,\sim_1), (S_2, \chi_2,\sim_2)$, and $a \in \Sigma^\pm$, we are able to define a syntactic successor relation $(S_1,\chi_1, \sim_1)\stackrel{a}{\longrightarrow}(S_2, \chi_2,\sim_2)$, which mimics the changes from $prof_{\alpha}(2i)$ to $prof_{\alpha}(2(i+1))$ by reading a word symbol $a$ in the position $2i+1$ of a data path. The details of the consistency condition and the successor relation are omitted due to the space limitation.

We are ready to construct the NRAG $\Bb$. 

There are $2|\Tt_\Aa|+1$ registers in $\Bb$, that is, 
\[r_1,\dots, r_{|\Tt_\Aa|}, r_{|\Tt_\Aa|+1}, \dots, r_{2|\Tt_\Aa|}.\]

Over a data path $\alpha=d_0 a_1 d_1 \dots a_n d_n$, $\Bb$ does the following.
\vspace*{-1mm}
\begin{itemize}
\item In each position $2i$ ($0 \le i \le n$), $\Bb$ guesses $\pi_i =(S_i,\chi_i,\sim_i) \in Prof_\Aa$ (where $\pi_i$ is supposed to be $prof_\alpha[2i]$). In addition, 
\begin{itemize}
\item if $i=0$, then $\pi_0=(S_0,\chi_0,\sim_0)$ is an initial profile, that is, for every $t \in \Tt_\Aa$ such that $\pred(\cur) \preceq t$ or $\pred_A(\cur) \preceq t$ for some $A \subseteq \Sigma^\pm$,  $t \nin S_0$,
\item if $i=n$, then $\pi_n=(S_n,\chi_n,\sim_n)$ is a final profile, that is, for every $t \in \Tt_\Aa$ such that $\suc(\cur) \preceq t$ or $\suc_A(\cur) \preceq t$ for some $A\subseteq \Sigma^\pm$, $t \nin S_n$. 
\end{itemize}
\vspace*{-2mm}
\item For every $i: 0 \le i \le n$, if 
\[\chi_i =\begin{array}{l}(b_{i,-m_{i,1}}, T_{i,-m_{i,1}}, b'_{i,-m_{i,1}},s_{i,-m_{i,1}}) \dots \\
(b_{i,-1}, T_{i,-1}, b'_{i,-1},s_{i,-1}) (b_{i,0}, T_{i,0},b'_{i,0},s_{i,0}) \\
(b_{i,1}, T_{i,1},b'_{i,1},s_{i,1}) \dots\\
(b_{i,m_{i,2}}, T_{i,m_{i,2}},b'_{i,m_{i,2}},s_{i,m_{i,2}})\end{array},\]
then after the position $2i$ is visited (that is, the reading head is in $2i+1$), for each $j: -m_{i,1} \le j \le m_{i,2}$, $\Bb$ stores in the register $r_{j+|\Tt_\Aa|}$ the data value corresponding to $T_{i,j}$. In particular, $\Bb$ stores the data value $d_i$ in $r_{|\Tt_\Aa|}$.
\vspace*{-1mm}
\item Over each pair of positions $2i$ and $2(i+1)$ (where $0 \le i < n$), $\Bb$ checks that $\pi_{i} \xrightarrow{a_{i+1}} \pi_{i+1}$. To do this, $\Bb$ copies (by guessing) data values between registers and guesses some data values for a few registers.
\vspace*{-1mm}
\item At the same time, $\Bb$ simulates the run of $\Aa$ as follows.
\begin{itemize}
\item If $\Aa$ makes a transition $(q,a_i,q')$ over $a_i$, then $\Bb$ checks that $b'_{i-1,0}=a_i$ and changes the state from $q$ to $q'$.
\item If $\Aa$ makes a transition $(q,c,q')$ over $d_i$, then $\Bb$ checks that $\pi_i$ satisfies $c$, verifies that $d_i$ is equal to the data value stored in $r_{j+|\Tt_\Aa|}$ for each $j: -m_1 \le j \le m_2$ such that there is $t \in T_j$ satisfying $\cur \sim_i t$ (in particular, $d_i$ should be equal to the data value in $r_{|\Tt_\Aa|}$), and changes the state from $q$ to $q'$.
\item $\Bb$ accepts if $\Aa$ accepts and a final profile is reached.
\end{itemize}
\end{itemize}
From the above construction, we know that in its states, $\Bb$ should record the states of $\Aa$ and the guessed profiles. Therefore, the number of states of $\Bb$ is polynomial over $|Q|$ and exponential over $|\Tt_\Aa|$. \qed
\end{proof}



\vspace*{-3mm}
\begin{proposition}\label{prop-nonempt-nrra}
The nonemptiness of NRRAs and DRRAs is PSPACE-complete.
\end{proposition}
\vspace*{-1mm}

By using a slight extension of the subset construction, we are able to show that NRRA can be determinized.

\vspace*{-2mm}
\begin{proposition}\label{prop-nrra-drra}
For every NRRA $\Aa$, there is an equivalent DRRA of exponential size.
\end{proposition}

\vspace*{-5mm}
\begin{corollary}
NRRAs are closed under all Boolean operations.
\end{corollary}

\vspace*{-5mm}
\begin{corollary}\label{cor-inc-nrra}
The language inclusion problem of NRRAs is PSPACE-complete.
\end{corollary}

\vspace*{-2mm}
\subsection{Two-way nondeterministic rigid register automata}

In this subsection, we will show that two-way nondeterministic rigid register automata are of the same expressibility as NRRA.

A \emph{two-way} nondeterministic rigid register automaton (2NRRA) $\Aa$ over $\Sigma^\pm$ is a tuple $(Q, \vdash,\dashv,\delta,I,F)$, where $Q,I,F$ are as those in the definition of NRRAs, $\vdash,  \dashv \nin \Sigma^\pm$ are respectively the left and right endmarkers, $\delta =\delta_w \cup \delta_d$ such that
\begin{itemize}
\item $\delta_w \subseteq Q \times (\Sigma^\pm \cup \{\vdash, \dashv\}) \times  Q \times \{+1,-1\}$ (where $+1,-1$ denote the direction of the head: ``right'' and ``left'') satisfies that for every transition $(q,\vdash,q',dir) \in \delta_w$ (resp. $(q,\dashv,q',dir) \in \delta_w$), it holds that $dir=+1$ (resp. $dir=-1$),
\item $\delta_d \subseteq Q \times \Cc_{rgd} \times Q \times \{+1,-1\}$.
\end{itemize}

Let $\alpha=d_0 a_1 d_1 \dots a_n d_n$ be a data path and $\Aa$ be a 2NRRA. A \emph{run} of $\Aa$ over $\alpha$  is a sequence 
\[(q_0,i_0) \theta_0 (q_1,i_1) \theta_1 \dots \theta_{m-1} (q_m,i_m)\]
such that $q_0 \in I$, $i_0=0$, $i_m=2n+2$,
\begin{itemize}
\item for every $j: 0 \le j < m$, if the symbol of $\vdash \alpha \dashv$ in the position $i_j$ is a word symbol $a \in \Sigma^\pm \cup \{\vdash,\dashv\}$, then there is $dir \in \{+1,-1\}$ such that $(q_j,a,q_{j+1},dir) \in \delta_w$, $\theta_j=a$, and $i_{j+1}=i_j+dir$, 
\item for every $j: 0 \le j \le m$, if the symbol of $\vdash \alpha \dashv$ in the position $i_j$ is a data value $d$, then there are $c \in \Cc_{rgd}$ and $dir \in \{+1,-1\}$ such that $(q_j,c,q_{j+1},dir) \in \delta_d$, $(\alpha,i_j-1) \models c$, $\theta_j=c$, and $i_{j+1}=i_j + dir$.
\end{itemize}
A run is accepting if $q_m \in F$. Note that a run of a 2NRRA over $\alpha$ starts at the left endmarker (position $0$) and stops at the right endmarker (position $2n+2$).

\vspace*{-2mm}
\begin{proposition}\label{prop-2nrra-nrra}
For every 2NRRA, there is an equivalent NRRA of exponential size.
\end{proposition}

\vspace*{-3mm}
\subsection{Rigid regular expressions with memory}

\emph{Rigid} regular expressions with memory (RREM) is defined by the following rules,
\[e:=\varepsilon \mid a \mid [c] \mid e \cup e \mid e \cdot e \mid e^+, \mbox{ where } c \in \Cc_{rgd}[\Sigma^\pm].\]

Let $e$ be a RREM, $\alpha=d_0 a_1 d_1\dots a_n d_n$ be a data path, and $i,j: 0 \le i \le j \le 2n$. The semantics of $e$ is defined by a relation $(\alpha,i) \vdash_e (\alpha,j)$ as follows.
\begin{itemize}
\item If $e=\varepsilon$, then $(\alpha,i) \vdash_{e} (\alpha,j)$ if $i=j$ and the symbol of $\alpha$ at position $i$ is a data value (thus $i$ is even).
\item If $e=a$, then $(\alpha,i) \vdash_{e} (\alpha,j)$ if $j=i+2$, the symbol of $\alpha$ at position $i+1$ is $a$.
\item If $e=[c]$, then $(\alpha,i) \vdash_{e} (\alpha,j)$ if $i=j$, the symbol of $\alpha$ at position $i$ is a data value (thus $i$ is even), and $(\alpha,i) \models c$.
\item The semantics for the rules $e_1 \cup e_2$, $e_1 \cdot e_2$ and $e^+_1$ are defined in a natural way and are omitted.
%
%
\end{itemize}

A data path $\alpha=d_0 a_1 d_1 \dots a_n d_n$ is accepted by a RREM $e$ if $(\alpha,0) \vdash_e (\alpha,2n)$. Let $\Ll(e)$ denote the set of data paths accepted by a RREM $e$.


\vspace*{-2mm}
\begin{proposition}\label{prop-nrra-rrem}
NRRAs and RREMs have the same expressive power.
\begin{itemize}
\item From a RREM $e$, a NRRA $\Aa_e$ can be constructed in LOGSPACE such that $\Ll(e)=\Ll(\Aa_e)$.
\item From a NRRA $\Aa$, a RREM $e_\Aa$ can be constructed in EXPTIME such that $\Ll(\Aa)=\Ll(e_\Aa)$. 
\end{itemize} 
\end{proposition}

\vspace*{-4mm}

\begin{corollary}
The nonemptiness problem of RREMs is PSPACE-complete.
\end{corollary}

\vspace*{-3mm}
\section{Rigid regular path queries with data}\label{sec-rrdpq}

A rigid regular path query with inverse and data (2RRDPQ) $\xi$ over the alphabet $\Sigma$ is a tuple $(x,L,y)$ where $L$ is a language of data paths defined by a NRRA or a RREM over $\Sigma^{\pm}$. 

Given a data graph $\dg=(V,E,\eta)$ and a RRDPQ $\xi=(x,L,y)$, the evaluation result of $\xi$ over $\dg$, denoted by $\xi(\dg)$, is the set of all pairs $(v,v')$ such that there is a semipath $\pi$ from $v$ to $v'$ in $\dg$ such that $\eta(\pi) \in L$.

\vspace*{-2mm}
\begin{proposition}\label{prop-complexity-2rrdpq}
The evaluation problem of 2RRDPQs is PSPACE-complete, and NLOGSPACE-complete in data complexity.
\end{proposition}
\vspace*{-1mm}

In the following we will show that every 2RDPQ can be turned into a 2RRDPQ, if data graphs are transformed in a natural way. Note that the transformation of data graphs presented in the following is \emph{localized} in the sense that for each node, a new node is added and connected to the node by edges with special labels, and the relationships between the nodes in the original data graph are not changed.

Let $\dg=(V,E,\eta)$ be a data graph over the alphabet $\Sigma$, $k \ge 1$, and $\{A_i \mid 1 \le i \le k\} \cap \Sigma=\emptyset$. Then the \emph{data-to-node $k$-transformation} of $\dg$, denoted by $\dg_{dn,k}=(V_{dn,k},E_{dn,k},\eta_{dn,k})$, is  defined as follows.
\begin{itemize}
\item $V_{dn,k}$ is obtained from $V$ by adding a new node $n_v$ for each node $v \in V$,
\item $E_{dn,k}$ is defined as the union of $E$ and the set of edges $(v,A_i,n_v)$ for every $v\in V$ and $i: 1 \le i \le k$,
\item for each $v \in V$, $\eta_{dn,k}(v)=\eta(v)$ and $\eta_{dn,k}(n_v)=\eta(v)$. 
\end{itemize}

The intuition of the above transformation is to copy the data value of each node $v$ to a new node connected to $v$ with $k$ edges. Note that the transformation does not change the edges between nodes in the original graph.

\vspace*{-2mm}
\begin{theorem}\label{thm-rdpq2rrdpq}
Let $k \ge 1$, $\xi=(x,L,y)$ be a 2RDPQ over the alphabet $\Sigma$ such that $L$ is given by a NRA or REM containing at most $k$-registers. Then a 2RRDPQ $\xi'=(x,L',y)$ over the alphabet $\Sigma^\pm \cup \{A_i, A_i^-\mid 1 \le i \le k\}$ can be constructed in polynomial time such that for every data graph $\dg=(V,E,\eta)$, $\xi(\dg)=\xi'(\dg_{dn,k})$.  
\end{theorem}
\vspace*{-1mm}

Note that in practice, the number $k$ in 2RDPQs are usually small, e.g. $k=1,2$, and can be assumed to be a constant. Then the above data-to-node transformation becomes query-independent. 

\vspace*{-2mm}
\section{Conjunctive rigid regular path queries with data}

Conjunctive 2RRDPQs (C2RRDPQ) can be defined similarly to C2RDPQs,  with 2RDPQs replaced by 2RRDPQs.





\vspace*{-2mm}
\begin{proposition}\label{prop-eval-c2rrdpq}
The evaluation of C2RRDPQs is PSPACE-complete, and NLOGSPACE-complete in data complexity.
\end{proposition}

\vspace*{-3mm}
\begin{theorem}\label{thm-crrdpq-containment}
The containment of C2RRDPQs is in 2EXPSPACE and EXPSPACE hard.
\end{theorem}
\vspace*{-1mm}

The rest of this section is devoted to the proof of Theorem \ref{thm-crrdpq-containment}. The proof is a nontrivial extension of that of the EXPSPACE-completeness result for C2RPQs in \cite{CGLV00} and is the most technical part of this paper.

\vspace*{-1mm}
\subsection{Canonical data graph}


Let $\xi:=Ans(\bar{z}) \leftarrow \bigwedge \limits_{1 \le i \le l} (y_{2i-1}, L_i, y_{2i})$ be  a C2RRDPQ, $\dg=(V,E,\eta)$ be a data graph, and $\nu: \{y_1,\dots,y_{2l}\} \rightarrow V$. Then $\dg$ is said to be \emph{$\nu$-canonical} for $\xi$ if 
\begin{itemize}
\item $\dg$ consists of $l$ simple semipaths $\pi_1,\dots,\pi_l$, one for each conjunct of $\xi$, such that only start and end nodes can be shared among different semipaths.
\item for every $i: 1 \le i \le l$, $\pi_i$ is a semipath from $\nu(y_{2i-1})$ to $\nu(y_{2i})$ such that $\eta(\pi_i)$ belongs to $L_i$.
\end{itemize}

It is easy to see that if $\dg$ is $\nu$-canonical for $\xi$, then $\nu(\bar{z})$ belongs to $\xi(\dg)$.

In the rest of this section, we assume that $\xi_1,\xi_2$ are two C2RRDPQs such that  
\begin{itemize}
\item $\xi_1$ and $\xi_2$ have the same set of distinguished variables, 
\item the set of non-distinguished variables of $\xi_1$ and $\xi_2$ are disjoint.
\end{itemize}
More specifically, for $i=1,2$, let 
\[\xi_i:=Ans(z_1,\dots,z_n) \leftarrow \bigwedge \limits_{1 \le j \le l_i} (y_{i,2j-1}, L_{i,j}, y_{i,2j})\]
such that $\{y_{1,1},\dots,y_{1,2l_1}\} \cap \{y_{2,1}, \dots,y_{2,2l_2}\}$ is equal to $\{z_1,\dots,z_n\}$.

Let $\dg=(V,E,\eta)$ be a $\nu$-canonical data graph for $\xi_1$. Then a mapping $\mu: \{y_{2,1}, \dots, y_{2,l_2}\} \rightarrow V$ is said to be a \emph{$(\xi_1,\dg,\nu)$-mapping} for $\xi_2$ if 
\begin{itemize}
\item for every $j: 1 \le j \le n$, $\nu(z_j)=\mu(z_j)$,
\item for every $j: 1 \le j \le l_2$, $(\mu(y_{2,2j-1}),\mu(y_{2,2j}))$ belongs to the evaluation result of $(y_{2,2j-1},L_{2,j},y_{2,2j})$ over $\dg$.
\end{itemize}
Note that the existence of a $(\xi_1,\dg,\nu)$-mapping for $\xi_2$ implies that $\nu(\bar{z}) \in \xi_2(\dg)$.

The following result can be shown in the same way as a corresponding result for C2RPQs (Theorem 2 in \cite{CGLV00}).

\vspace*{-2mm}
\begin{proposition}
Let $\xi_1,\xi_2$ be two C2RRDPQs. Then $\xi_1 \not \subseteq \xi_2$ iff there are a data graph $\dg$ and a mapping $\nu$ from the variables in $\xi_1$ to the nodes in $\dg$ such that 
\begin{itemize}
\item $\dg$ is $\nu$-canonical for $\xi_1$, 
\item and there are no $(\xi_1,\dg,\nu)$-mappings for $\xi_2$.
\end{itemize}
\end{proposition}

\vspace*{-2mm}
\subsection{Evaluating 2RRDPQs over canonical data graphs}

Let $\dg=(V,E,\eta)$ be a $\nu$-canonical data graph for $\xi_1$ and $\xi=(x,L,y)$ be a 2RRDPQ such that $L$ is defined by a NRRA $\Aa=(Q,\delta,I,F)$ over $\Sigma^{\pm}$. Then $\dg$ consists of $l_1$-simple semipaths $\pi_1,\dots,\pi_{l_1}$ such that for every $j: 1 \le j \le l_1$, $\pi_j$ is a semipath from $\nu(y_{1,2j-1})$ to $\nu(y_{1,2j})$, and $\eta(\pi_j) \in L_{1,j}$. Our goal is to evaluate $\xi$ over $\dg$.

We use a similar idea to the evaluation of 2RPQs over canonical graphs in \cite{CGLV00}: the data graph $\dg$ is first encoded into a data path $\alpha_\dg$, then a 2NRRA $\Aa_\xi$ is constructed from $\xi$ and $\xi_1$ so that  $\xi(\dg)$ is nonempty iff   $\vdash \alpha_\dg \dashv$ is accepted by $\Aa_\xi$.

For every $i:1 \le i \le l_1$, let $\ren_i$ denote the renaming function that maps each $a \in \Sigma^{\pm}$ to $(a,i)$. For a data path $\alpha$, let $\ren_i(\alpha)$ denote the data path obtained from $\alpha$ by replacing each $a \in \Sigma^{\pm}$ with $\ren_i(a)$.

Let  $\Sigma_{\xi_1}=\{\#\} \cup \bigcup \limits_{1 \le j \le l_1} (\Sigma^{\pm} \times \{j\} \cup \{\$_{2j-1},\$_{2j}\})$. We represent $\dg$ as a data path $\alpha_\dg$ over the alphabet $\Sigma_{\xi_1}$ as follows. 
\[\alpha_\dg:=\begin{array}{c}d_1 \$_1 \ren_1(\eta(\pi_1)) \$_2 d_2 \#  
d_3 \$_3 \ren_2(\eta(\pi_2)) \$_4 d_4  \#\\
\dots d_{2l_1-1} \$_{2l_1-1} \ren_{l_1}(\eta(\pi_{l_1})) \$_{2l_1} d_{2l_1}\end{array},\]
where $d_1,d_2,\dots,d_{2l_1}$ are data values from $\Dd$ not occurring in $\dg$ such that $d_i=d_j$ iff $\nu(y_{1,i}) = \nu(y_{1,j})$. Intuitively, for each $j: 1 \le j \le l_1$, the $j$-th semipath $\pi_j$ is represented by a data subpath $\alpha_{\dg,\pi_j}$ in $\alpha_\dg$, where $\alpha_{\dg,\pi_j}=d_{2j-1} \$_{2j-1} \ren_j(\eta(\pi_j)) \$_{2j} d_{2j}$, and the symbol $\#$ is used to separate those data subpaths. It is easy to observe that for every pair $(\pi_j,v)$ such that $v$ is a node in $\pi_j$, there is a unique position in $\alpha_\dg$ corresponding to $(\pi_j,v)$, denoted by $p_{\alpha_\dg}(\pi_j,v)$. For instance, if $v=\nu(y_{1,2j})=\nu(y_{1,2j'-1})$, then $p_{\alpha_\dg}(\pi_j,v)$ is the position immediately before the symbol $\$_{2j}$ and $p_{\alpha_\dg}(\pi_{j'},v)$ is the position immediately after $\$_{2j'-1}$ in  $\alpha_{\dg}$.

For the simplicity of presentations, we assume that for every $j: 1 \le j \le l_1$, $\pi_j$ contains at least two edges. All the proofs in the rest of this section can be easily adapted to deal with the situation that there is $j: 1 \le j \le l_1$ such that $\pi_j$ contains at most one edge.

Let $\pi'=v_0 a_1 v_1 \dots v_{\ell-1} a_\ell v_\ell$ be a semipath in $\dg$ (since $\pi'$ is an arbitrary semipath in $\dg$, it may start or end in the middle of $\pi_1,\dots,\pi_{l_1}$). Because $\alpha_\dg$ is an encoding of the data graph $\dg$ and $\pi'$ is a semipath in $\dg$,  there is also an encoding of $\pi'$ in $\alpha_\dg$. We call this encoding as the \emph{trace} of $\pi'$ in $\alpha_\dg$, denoted by $trc_{\alpha_\dg}(\pi')$. A formal definition of $trc_{\alpha_\dg}(\pi')$ will be given later.

The intuition of  the 2NRRA $\Aa_\xi$ is that for every semipath $\pi'$ of $\dg$ and every run of $\Aa$ over $\eta(\pi')$, $\Aa_\xi$ goes through the trace of $\pi'$ in $\alpha_\dg$ to simulate the run of $\Aa$ over $\eta(\pi')$. 

\vspace*{-2mm}
\begin{theorem}\label{thm-rrdpq-auto}
Let $\dg$ be a $\nu$-canonical data graph for $\xi_1$, $\xi$ be a 2RRDPQ. Then a 2NRRA $\Aa_\xi$ can be constructed from $\xi$ and $\xi_1$ such that $\xi(\dg)$ is nonempty iff $\Aa_{\xi}$ accepts $\vdash \alpha_\dg \dashv$. 
\end{theorem}
\vspace*{-1mm}

In the following, before giving a proof for Theorem \ref{thm-rrdpq-auto}, we first give the definition of traces of semipaths of $\dg$ in $\alpha_\dg$, then state and prove an important lemma.

Let $\pi'=v_0 a_1 v_1 \dots v_{\ell-1} a_\ell v_\ell$ be a semipath in $\dg$. The \emph{$\overline{\pi}$-unraveling} (where $\overline{\pi}=(\pi_1,\dots,\pi_{l_1})$) of $\pi'$, denoted by $urv_{\overline{\pi}}(\pi')$, is defined as the sequence $\pi'_0 \# \pi'_1 \#\dots\# \pi'_r$ satisfying the following conditions: There are $i_0,\dots, i_{r+1}$ such that 
\begin{itemize}
\item $0=i_0 < i_1 < \dots < i_r< i_{r+1}=\ell$,  
\item for every $0 \le s \le r$, $\pi'_s = v_{i_s} a_{i_s+1} v_{i_s+1}\dots a_{i_{s+1}} v_{i_{s+1}}$,  
\item for every $s: 0 \le s \le r$, there is $j_s: 1 \le j_s \le l_1$ such that all the edges on $\pi'_s$ belong to $\pi_{j_s}$, 
\item and for every $s: 1 \le s \le r$, either $j_s \neq j_{s-1}$, or $j_s=j_{s-1}$ and one of the following conditions holds,
\begin{itemize}
\item the last edge of $\pi'_{s-1}$ is the first edge of $\pi_{j_{s}}$ and the first edge of $\pi'_s$ is the last edge of $\pi_{j_s}$,
\item the last edge of $\pi'_{s-1}$ is the last edge of $\pi_{j_{s}}$ and the first edge of $\pi'_s$ is the first edge of $\pi_{j_s}$.
\end{itemize}
\end{itemize}
\vspace*{-2mm}

The last two conditions above correspond to the situation that the two endpoints of $\pi_{j_s}$ are in fact the same node and a semipath can jump from the first (resp. last) edge to the last (resp. first) edge of $\pi_{j_s}$.

For a semipath $\pi'$ in $\dg$, define $trc_{\alpha_{\dg}}(\pi')$, the trace of $\pi'$ in $\alpha_\dg$, as  $trc_{\alpha_\dg}(\pi'_0) \ \#  \dots \  \# \ trc_{\alpha_\dg}(\pi'_r)$, where $\pi'_0 \# \dots \# \pi'_r$ is the $\overline{\pi}$-unraveling of $\pi'$, and for every $s: 0 \le s \le r$, 
\[trc_{\alpha_\dg}(\pi'_s):=\begin{array}{c}p_{\alpha_\dg}(\pi_{j_s},v_{i_s}) \ a_{i_s+1}\ p_{\alpha_\dg}(\pi_{j_s},v_{i_s+1}) \\
\dots \ a_{i_{s+1}}\ p_{\alpha_\dg}(\pi_{j_s},v_{i_{s+1}})\end{array}.\]

Note that although $urv_{\overline{\pi}}(\pi')$ and $trc_{\alpha_{\dg}}(\pi')$ are not data paths, they are of a similar structure, that is, nodes and position indices respectively separated by word symbols.

For briefness, later on, when $\alpha_\dg$ is obvious from the context, we abbreviate $trc_{\alpha_\dg}(\pi')$ as $trc(\pi')$.

It is easy to see that a run of $\Aa$ over $\eta(\pi')$ for a semipath $\pi'$ in $\dg$ can be transformed into a run of a NRRA $\Aa'$ over $\eta(urv_{\overline{\pi}}(\pi'))$, if the interpretation of position terms over $\eta(urv_{\overline{\pi}}(\pi'))$ is adjusted to jump over the additional $\#$ symbols as follows.

Since $\eta(urv_{\overline{\pi}}(\pi'))=\eta(\pi'_0) \# \dots \# \eta(\pi'_r)$, it follows that $\eta(urv_{\overline{\pi}}(\pi'))=d_0 b_1 d_1 \dots b_{\ell+2r} d_{\ell+2r}$ for $d_0,\dots,d_{\ell+2r} \in \Dd$ and $b_1,\dots,b_{\ell+2r} \in \Sigma^\pm \cup \{\#\}$.
For a position term $t \in \Tt_p[\Sigma^\pm]$ and a position $2i: 0 \le i \le \ell+2r$ on $\eta(urv_{\overline{\pi}}(\pi'))$, define the \emph{adjusted position} represented by $t$ over $\eta(urv_{\overline{\pi}}(\pi'))$ and $2i$, denoted by $t^{adj}_{urv_{\overline{\pi}}(\pi')}[2i]$, similarly to the semantics of position terms, with the following adjustments for the rules $\suc(t_1)$ and $\pred(t_1)$. In the following, we only present the adjustments for $\suc(t_1)$, and the adjustments for $\pred(t_1)$ are symmetric.
If $(t_1)^{adj}_{urv_{\overline{\pi}}(\pi')}[2i]=\bot$ or $(t_1)^{adj}_{urv_{\overline{\pi}}(\pi')}[2i]=2(\ell+2r)$, then $(\suc(t_1))^{adj}_{urv_{\overline{\pi}}(\pi')}[2i]=\bot$; otherwise, 
\begin{itemize}
\item if $(t_1)^{adj}_{urv_{\overline{\pi}}(\pi')}[2i]$ is not a position immediately before $\#$, then 
\[(\suc(t_1))^{adj}_{urv_{\overline{\pi}}(\pi')}[2i]=(t_1)^{adj}_{urv_{\overline{\pi}}(\pi')}[2i]+2,\] 
\item otherwise, 
\[(\suc(t_1))^{adj}_{urv_{\overline{\pi}}(\pi')}[2i]=(t_1)^{adj}_{urv_{\overline{\pi}}(\pi')}[2i]+4.\]
\end{itemize}
%

\vspace*{-3mm}
\begin{lemma}\label{lem-pt-pt}
Suppose $\pi'=v_0 a_1 v_1 \dots a_\ell v_\ell$ is a semipath in $\dg$ such that $urv_{\overline{\pi}}(\pi')=\pi'_0 \# \dots \# \pi'_r$ and $trc(\pi')=p_0 b_1 p_1 \dots b_{\ell+2r} p_{\ell+2r}$ (where $b_1,\dots,b_{\ell+2r} \in \Sigma^\pm \cup \{\#\}$). Then for every $i: 0 \le i \le \ell+2r$, there exists a function $pos_i \in (\Tt_p[\Sigma_{\xi_1}] \cup \{\bot\})^{\Tt_\Aa}$ such that for every $t \in \Tt_\Aa$,  $pos_i(t) =\bot$ iff $t^{adj}_{urv_{\overline{\pi}}(\pi')}[2i]=\bot$; moreover, if $pos_i(t) \neq \bot$ and $t^{adj}_{urv_{\overline{\pi}}(\pi')}[2i]=2i'$, then $(pos_i(t))_{\alpha_\dg}[p_i]=p_{i'}$.
\end{lemma}
\vspace*{-1mm}

Lemma \ref{lem-pt-pt} establishes a connection between the position terms in $\Tt_p[\Sigma]$ interpreted over $\eta(urv_{\overline{\pi}}(\pi'))$ and the position terms in $\Tt_p[\Sigma_{\xi_1}]$ interpreted over $\alpha_\dg$. With this connection, a 2NRRA $\Bb$ can be constructed such that each run of $\Aa'$ over $\eta(urv_{\overline{\pi}}(\pi'))$ can be simulated by a run of $\Bb$ over $trc(\pi')$ in $\alpha_\dg$.


\begin{proof} (Theorem \ref{thm-rrdpq-auto})

Let $\pi'$ be a path in $\dg$, the $\overline{\pi}$-unraveling of $\pi$ be $\pi'_0 \dots \pi'_r$. In addition, for every $s: 0 \le s \le r$, all the edges on $\pi'_s=v_{i_s} a_{i_s+1} v_{i_s+1} \dots v_{i_{s+1}}$ belong to $\pi_{j_s}$.


Our goal is to construct a 2NRRA $\Bb$ over $\alpha_\dg$ to simulate the runs of $\Aa'$ over $\eta(urv_{\overline{\pi}}(\pi'))$.

Similarly to the construction of NRAGs from NRRAs in the proof of Theorem \ref{thm-nrra-nrag}, the 2NRRA $\Bb$ goes through $trc(\pi')$ in $\alpha_\dg$ and guesses the profile of the current position of $\eta(urv_{\overline{\pi}}(\pi'))$, in order to simulate $\Aa'$ over $\eta(urv_{\overline{\pi}}(\pi'))$. The difference is that instead of storing and guessing the data values, $\Bb$ records and guesses position terms from $\Tt_p[\Sigma_{\xi_1}]$ (interpreted over $\alpha_\dg$) for position terms occurring in the profile of the current position in $\eta(urv_{\overline{\pi}}(\pi'))$. The most technical part of the construction is how to guarantee the consistency of the guessed position terms from $\Tt_p[\Sigma_{\xi_1}]$ and how to update them during the simulation. Since the details of the consistency conditions and the updating of the guessed position terms are rather tedious, they are omitted due to the space limitation.

From the above description, we know that in its states, $\Bb$ should record the states of $\Aa'$, the guessed profiles, and the guessed position terms from $\Tt_p[\Sigma_{\xi_1}]$. Because both the number of profiles and the number of possible guesses for the position terms from $\Tt_p[\Sigma_{\xi_1}]$ are exponential over $|\Tt_\Aa|$, it follows that the number of states of $\Bb$ is polynomial over $|Q|$ and exponential over $|\Tt_\Aa|$. 
\end{proof}

\vspace*{-3mm}

\subsection{Checking the non-containment}

We will construct a NRRA $\Aa'=(Q',\delta',I',F')$ to check the non-containment of $\xi_1$ over $\xi_2$ as follows.

\begin{enumerate}
\item Construct a NRRA $\Aa'_1$ which reads a data path $\alpha$ over the alphabet $\Sigma_{\xi_1}$ and verifies that $\alpha$ encodes a $\nu$-canonical data graph $\dg$ for $\xi_1$. In particular,  for every 2RRDPQ $(y_{1,2j-1},L_{1,j},y_{1,2j})$, $\Aa'_1$ checks that the $j$-th block of $\alpha$ encodes a data path over the alphabet $\Sigma^\pm$ belonging to $L_{1,j}$.
\item Construct a NRRA $\Aa'_2$ verifying that there are no $(\xi_1, \dg,\nu)$-mappings for $\xi_2$ as follows.
\begin{enumerate}
\item Construct a 2NRRA $\Bb_1$ to verify a $(\xi_1, \dg,\nu)$-mapping for $\xi_2$ over $\alpha_\dg$ annotated with subsets of $\{y_{2,1},\dots,y_{2,l_2}\}$. The intention is that the annotations encode an assignment of nodes in $\dg$ to the variables from $\{y_{2,1},\dots,y_{2,l_2}\}$. The alphabet of $\Bb_1$ is $\Sigma^e_{\xi_1}=\Sigma_{\xi_1} \times 2^{\{y_{2,1},\dots,y_{2,l_2}\}}$. If the word symbol immediately before a position $2i$ of the annotated $\alpha_\dg$ is $(a',Z)$, then this means that each variable in $Z$ is assigned to the node of $\dg$ represented by the position $2i$. Some consistency constraints for these annotations, e.g. the annotations in two distinct positions are disjoint, should be checked. To check the 2RRDPQs $(y_{2,2j-1}, L_{2,j}, y_{2,2j})$ of $\xi_2$ over the annotated $\alpha_\dg$, the construction in the proof of Theorem \ref{thm-rrdpq-auto} is used. Note that since all the rigid data constraints in the RRDPQs of $\xi_2$ are independent from the annotations, we are able to assume that for every $\suc_{B}$ or $\pred_B$ occurring in $\Bb_1$, there is $A \subseteq \Sigma_{\xi_1}$ such that $B=A \times 2^{\{y_{2,1},\dots,y_{2,l_2}\}}$.
\item Transform $\Bb_1$ into an equivalent NRRA $\Bb_2$ (cf. Proposition \ref{prop-2nrra-nrra}). 
\item Let $prj: \Sigma^e_{\xi_1} \rightarrow \Sigma_{\xi_1}$ such that $prj((a',Z))=a'$. Construct $\Bb_3=prj(\Bb_2)$. From the assumption above, we know that $\Bb_1$ and $\Bb_2$ are position-invariant under $prj$. Then from Proposition \ref{prop-nrra-pos-inv}, we deduce that $\Ll(\Bb_3)=\Ll(prj(\Bb_2))=prj(\Ll(\Bb_2))$. So the NRRA $\Bb_3$ guesses and verifies a $(\xi_1, \dg,\nu)$-mapping for $\xi_2$. 
\item Determinize and complement $\Bb_3$ to get $\Aa'_2$ (cf. Proposition \ref{prop-nrra-drra}).
\end{enumerate}
\item $\Aa'$ is the intersection of $\Aa'_1$ and $\Aa'_2$.
\end{enumerate}
There is a final remark for the above construction: 
As pointed out in \cite{CGLV00}, letter projections are only meaningful for one-way automata. This explains why we need go from the 2NRRA $\Bb_1$ to the NRRA $\Bb_2$ before applying the letter projection $prj$.


\noindent {\bf The complexity analysis}.

The size of $\Aa'_1$ is polynomial over the size of $\xi_1$. From Theorem \ref{thm-rrdpq-auto}, the size of $\Bb_1$ is exponential over the size of $\xi_2$.  From Proposition \ref{prop-2nrra-nrra}, the size of $\Bb_2$ is exponential over the size of $\Bb_1$. The size of $\Bb_3$ is the same as the size of $\Bb_2$. From Proposition \ref{prop-nrra-drra}, the size of $\Aa'_2$ is exponential over the size of $\Bb_3$. Therefore, the size of $\Aa'_2$ is triple-exponential over the size of $\xi_2$.

To check the nonemptiness of $\Aa'_1\cap \Aa'_2$, we can guess ``on the fly'' an accepting run of $\Aa'_1 \cap \Aa'_2$ in double exponential space. From Savitch's theorem, we deduce that the containment of C2RRDPQs is in 2EXPSPACE. 

On the other hand, the containment of C2RRDPQs is EXPSPACE-hard since this is already the case for C2RPQs (\cite{CGLV00}).

\vspace*{-1mm}

\section{Conclusion}

In this paper, a novel approach to extend 2RPQs with data value comparisons, called rigid regular path queries with inverse and data (2RRDPQs), was proposed. 2RRDPQs rely on nondeterministic rigid register automata (NRRA), also introduced in this paper. We demonstrated the robustness of NRRAs by showing that NRRAs can be determinized and the two-way NRRAs are expressively equivalent to NRRAs. We then argued that 2RRDPQs achieve a good balance between the expressibility and computational properties. On the one hand, we showed that every 2RDPQ  can be turned into a 2RRDPQ if a localized transformation is applied to graph databases. On the other hand, we proved that 2RRDPQs enjoy nice computational properties, as witnessed by the decidability (as a matter of fact, 2EXPSPACE) of the containment problem of 2RRDPQs and conjunctive 2RRDPQs (C2RRDPQ), The proof for the 2EXPSPACE result of the containment problem of C2RRDPQs is the most technical part of this paper and can be seen as the main result of this paper.

There are several natural directions for future work. One direction is to investigate whether the evaluation and containment problem of acyclic C2RRDPQs have a lower complexity. Another direction is to investigate nested rigid regular expressions with memory.

\vspace*{-1mm}


%
%
%


\bibliographystyle{plain}

\bibliography{gdb-data}

\newpage

\onecolumn

\begin{appendix}



%
%
%
%
%
%

\section{Proofs in Section 3.1}

\noindent{\bf Proposition \ref{prop-sat-rdc}}
{\it The satisfiability problem of rigid data constraints is NP-complete.}

\begin{proof}
Lower bound: By an easy reduction from the satisfiability of Boolean formulas.

Upper bound: Let $c$ be a rigid data constraint over $\Sigma^\pm$. 

Let $\Tt_c$ denote the minimal set of position terms satisfying the following conditions.
\begin{itemize}
\item for every $t_1 \sim t_2$ or $t_1 \nsim t_2$ occurring in $c$, $t_1,t_2 \in \Tt_c$,
\item for every $t \in \Tt_c$ and $t' \in \Cc_{rgd}[\Sigma^\pm]$ such that $t' \preceq t$, we have $t', t[t' \backslash \cur] \in \Tt_c$.
\end{itemize}

Similar to the construction of a NRAGs from NRRAs in the proof of Theorem \ref{thm-nrra-nrag}, we can define concept of profiles with respect to $\Tt_c$. More specifically, a profile is a triple $(S,\chi,\sim)$ such that $S \subseteq \Tt_c$, $\sim$ is an equivalence relation on $S$, and 
\[\chi =\begin{array}{l}(b_{-m_1}, T_{-m_1}, b'_{-m_1}, s_{-m_1}) \dots (b_{-1}, T_{-1}, b'_{-1},s_{-1})
 (b_0, T_0,b'_0,s_{0})  (b_1, T_1,b'_1,s_1) \dots(b_{m_2}, T_{m_2},b'_{m_2},s_{m_2})\end{array}.\]
In addition, some consistency conditions can be defined such that $c$ is satisfiable iff there is a consistent profile $(S,\chi,\sim)$ with $c \in T_0$ in $\chi$.

Since the size of a profile is polynomial over that of $c$, a profile $(S,\chi,\sim)$ can be guessed and the consistency condition as well as $c \in T_0$ can be checked in polynomial time. Therefore, the satisfiability of rigid data constraints is in NP.
\end{proof}

\section{Proofs in Section 3.2} 

\noindent {\bf Proposition \ref{prop-expr-nra-nrra}}
{\it NRAs and NRRAs are expressively incomparable.
}

\begin{proof}
The data language ``there are two distinct positions with the same data value'' is definable in NRAs, but not in NRRAs.

On the other hand, the data language ``the sequence of word symbols belongs to $ab^\ast a$ and the last data value does not occur elsewhere'' is definable in NRRAs, but not in NRAs.
\end{proof}

\smallskip

\noindent{\bf Proposition \ref{prop-let-proj}}
{\it The class of languages definable by NRRAs are not closed under letter projections.
}

\begin{proof}
Let $\Sigma=\{(a,0),(a,1)\}$, $\Gamma=\{a\}$, and $prj$ be a letter projection from $\Sigma$ to $\Gamma$ such that $prj((a,0))=prj((a,1))=a$.

Let  $L$ be the data language ``there are \emph{exactly} two distinct positions labeled by $(a,1)$ and the data values before these two positions are the same''. Then $prj(L)$ is the data language ``there are two distinct positions with the same data value''.

It is easy to see that $L$ can be defined by a NRRA $\Aa$. On the other hand, $prj(L)$ is not definable by a NRRA.

We would like to remark that over the alphabet $\{a\}$, the position terms $\suc_A(t)$ (resp. $\pred_A(t)$) in $prj(\Aa)$ are equal to $\suc(t)$ (resp. $\pred(t)$). Therefore, $prj(\Aa)$ does not define $prj(L)$.
\end{proof}

\smallskip

\noindent {\bf Proposition \ref{prop-nrra-pos-inv}}
{\it Suppose $\Aa$ is a NRRA over $\Sigma^\pm$ and $prj$ is a letter projection from $\Sigma^\pm$ to $\Gamma$. If $\Aa$ is position-invariant under $prj$, then $\Ll(prj(\Aa))=prj(\Ll(\Aa))$.}

\begin{proof}

Suppose $\Aa$ is a NRRA over $\Sigma^\pm$ and $prj$ is a letter projection from $\Sigma^\pm$ to $\Gamma$ such that $\Aa$ is position-invariant under $prj$.

For every position term $t \in \Tt_\Aa$, define $prj(t)$ as the position term obtained from $t$ by replacing every occurrence of $\suc_A$ with $\suc_{prj(A)}$. In addition, for every $c \in \Cc_\Aa$, define $prj(c)$ as the rigid data constraint obtained from $c$ by replacing every position term $t$ with $prj(t)$. 

We first prove the following claim.

\medskip

\noindent {\bf Claim}. Let $\alpha=d_0 a_1 d_1 \dots a_n  d_n $ be a data path, $c \in \Cc_\Aa$ and $i: 0 \le i \le n$. Then $(\alpha,2i) \models c$ iff $(prj(\alpha),2i) \models prj(c)$. 

\begin{proof}
It is sufficient to prove that for every $t \in \Tt_\Aa$ and $i: 0 \le i \le n$, $t_\alpha[2i]=(prj(t))_{prj(\alpha)}[2i]$.

This result can be proved by an induction on the structure of the position terms. In the following, we take $t=\suc_A(\cur)$ as an example to illustrate the proof.

Suppose $(\suc_A(\cur))_{\alpha}[2i]=2j$ for some $j: i < j$. Then the first occurrence of word symbols from $A$ in $\alpha$ after the position $2i$ is in the position $2j$. It follows that the first occurrence of word symbols from $prj(A)$ in $prj(\alpha)$ after the position $2i$ is in the position $2j$. Otherwise, there is $j': i < j' < j$ such that a word symbol from $prj(A)$ occurs in the position $2j'$ of $prj(\alpha)$. From the fact that $\Aa$ is position-invariant under $prj$, we know that $A=prj^{-1}(prj(A))$. Thus, a word symbol from $A$ occurs in the position $2j'< 2j$ of $\alpha$, a contradiction. Therefore, $(\suc_{prj(A)}(\cur))_{prj(\alpha)}[2i]=2j$.

Suppose $(\suc_{prj(A)}(\cur))_{prj(\alpha)}[2i]=2j$ for some $j: i < j$. Then the first occurrence of word symbols from $prj(A)$ in $prj(\alpha)$ after the position $2i$ is in the position $2j$. It follows that the first occurrence of word symbols from $A$ in $\alpha$ after the position $2i$ is in the position $2j$. Otherwise, there is $j': i < j' < j$ such that a word symbol from $A$ occurs in the position $2j'$ of $\alpha$. Thus, a word symbol from $prj(A)$ occurs in the position $2j'<2j$ of $prj(\alpha)$, a contradiction. Therefore, $(\suc_{A}(\cur))_{\alpha}[2i]=2j$.  
\end{proof}

$\Ll(prj(\Aa)) \subseteq prj(\Ll(\Aa))$:

Suppose $\beta = d_0 \gamma_1 d_1 \dots \gamma_n d_n \in \Ll(prj(\Aa))$. 

Then there is an accepting run of $prj(\Aa)$ over $\beta$, say $\rho=q_0 c_0 q_1 \gamma_1 q_2 c_1 \dots q_{2n-1} \gamma_n q_{2n} c_n q_{2n+1}$.

From the definition of $prj(\Aa)$, we know that 
\begin{itemize}
\item for every $i: 1 \le i \le n$, there is $a_i \in \Sigma^\pm$ such that $prj(a_i)=\gamma_i$ and $(q_{2i-1},a_i,q_{2i}) \in \delta_w$,
\item for every $i: 0 \le i \le n$, there is $c'_i \in \Cc_\Aa$ such that $(q_{2i}, c'_i, q_{2i+1}) \in \delta_d$ and $c_i=prj(c'_i)$.
\end{itemize}
Let $\alpha=d_0 a_1 d_1 \dots a_n d_n$. Then $\beta=prj(\alpha)$.

From the claim, we know that for every $i: 0 \le i \le n$, $(\alpha,2i) \models c'_i$ iff $(prj(\alpha),2i) \models c_i$. 

Therefore, $q_0 c'_0 q_1 a_1 q_2 \dots q_{2n-1} a_n q_{2n} c'_n q_{2n+1}$ is an accepting run of $\Aa$ over $\alpha$. We conclude that $\alpha \in \Ll(\Aa)$ and $\beta = prj(\alpha) \in prj(\Ll(\Aa))$.

\medskip

$prj(\Ll(\Aa)) \subseteq \Ll(prj(\Aa))$: 

Let $\beta=d_0 \gamma_1 d_1 \dots \gamma_n d_n \in prj(\Ll(\Aa))$. Then there is $\alpha = d_0 a_1 d_1 \dots a_n d_n  \in \Ll(\Aa)$ such that $\beta=prj(\alpha)$.  So there is an accepting run of $\Aa$ over $\alpha$, say $\rho=q_0 c_0 q_1 a_1 q_2 c_1 \dots q_{2n-1} a_n q_{2n} c_n q_{2n+1}$. 

From the claim, we know that for every $i: 0 \le i \le n$, $(\alpha,2i) \models c_i$ iff $(prj(\alpha),2i) \models prj(c_i)$. Therefore, $q_0 \ prj(c_0)\ q_1 \ prj(a_1)\ q_2 \ prj(c_1) \dots q_{2n-1}\ prj(a_n)\ q_{2n}\ prj(c_n)\ q_{2n+1}$ is an accepting run of $prj(\Aa)$ over $prj(\alpha)=\beta$. It follows that $\beta \in \Ll(prj(\Aa))$.\qed
\end{proof}

\smallskip

\noindent {Proposition \ref{prop-nonempt-nrag}}
{\it The nonemptiness problem of NRAGs is PSPACE-complete.}

\begin{proof}
The upper bound:  

Let $\Aa=(Q,k,\delta,I,F)$ be a NRAG. Then similar to NRAs (\cite{DL09, LV12a}), a NFA $\Bb=(Q',\delta',I',F')$ can be constructed such that $\Ll(\Aa)$ is nonempty iff $\Ll(\Bb)$ is nonempty, and $|Q'|$ is polynomial over $|Q|$ and exponential over $k$. To decide the nonemptiness of $\Bb$, an accepting run of $\Bb$ can be guessed nondeterministically in polynomial space. From Savitch's theorem, we know that the nonemptiness of $\Aa$ can be decided in PSPACE.

The lower bound: Follows from that of NRAs.
\end{proof}

\smallskip

\noindent{\bf Lemma \ref{lem-nrra-pos-bnd}}.
{\it Let $\Aa=(Q,\delta,I,F)$ be a NRRA over the alphabet $\Sigma^\pm$ and $\alpha=d_0 a_1 d_1 \dots a_n d_n$ be a data path. Then for every run $\rho$ of $\Aa$ over $\alpha$ and every $i: 0 \le i \le n$, $Pos^f_{\rho}[\alpha,2i] \cup Pos^p_{\rho}[\alpha,2i] \subseteq \{t_\alpha[2i] \mid t \in \Tt_\Aa\}$.}

\begin{proof}
Let $\Aa=(Q,\delta,I,F)$ be a NRRA, $\alpha=d_0 a_1 d_1 \dots a_n d_n$ be a data path,  $\rho=q_0 c_0 q_1 a_1 q_2 \dots q_{2n-1} a_n q_{2n} c_n q_{2n+1}$
be a run of $\Aa$ over $\alpha$, and $i: 0 \le i \le n$.

Let $t_\alpha[2j] \in Pos^f_\rho[\alpha,2i]$ such that $j \le i$, $t$ occurs in $c_j$, $t_\alpha[2j] \neq \bot$ and $t_\alpha[2j]> 2i$. Then there is $t' \in \Tt_p[\Sigma^\pm]$ such that $t' \preceq t$, $t'_\alpha[2j] \le 2i$, and for every $t'': t' \prec t'' \preceq t$, $t''_\alpha[2j]>2i$.  It follows that $\suc(t') \preceq t$ or $\suc_A(t') \preceq t$ for some $A \subseteq \Sigma^\pm$. 
\begin{itemize}
\item If $\suc(t') \preceq t$, then $t'_{\alpha}[2j]=2i$, since $(\suc(t'))_\alpha[2j]=t'_\alpha[2j]+1> 2i$ and $t'_\alpha[2j] \le 2i$. Thus, $t_\alpha[2j]=(t[t' \backslash \cur])_\alpha[t'_\alpha[2j]]=(t[t' \backslash \cur])_\alpha[2i]$.
\item If $\suc_A(t') \preceq t$, then $t'_\alpha[2j] \le 2i$ and $(\suc_A(t'))_\alpha[2j] > 2i$. It follows that $a_{((\suc_A(t'))_\alpha[2j])/2} \in A$, and for every $j': i < j' < ((\suc_A(t'))_\alpha[2j])/2$,  $a_{j'} \nin A$. From this, it is deduced that $(\suc_A(\cur))_\alpha[2i]=(\suc_A(t'))_\alpha[2j]=(\suc_A(\cur))_\alpha[t'_\alpha[2j]]$. Therefore, $t'_\alpha[2j]=2i$, and $t_\alpha[2j]=(t[t'  \backslash \cur])_\alpha[t'_\alpha[2j]]=(t[t'  \backslash \cur])_\alpha[2i]$.    
\end{itemize}

From the above argument, it follows that $Pos^f_\rho[\alpha,2i] \subseteq \{t_\alpha[2i] \mid t \in \Tt_\Aa\}$. Similarly, we can show that $Pos^p_{\rho}[\alpha,2i] \subseteq \{t_\alpha[2i] \mid t \in \Tt_\Aa\}$. 
\end{proof}

\smallskip

\noindent{\bf Theorem \ref{thm-nrra-nrag}}.
{\it From a NRRA $\Aa=(Q,\delta,I,F)$, an equivalent NRAG $\Bb=(Q',k,\delta',I',F')$ can be constructed such that $|Q'|$ is polynomial over $|Q|$ and exponential over $|\Tt_\Aa|$ and $k$ is polynomial over $|\Tt_\Aa|$.}

\begin{proof}
Let $\Aa=(Q,\delta,I,F)$ be a NRRA. In the following, we will construct a NRAG $\Bb$ to simulate $\Aa$.

We first give an intuitive description of the construction. Let $\rho$ be a run of $\Aa$ over a data path $\alpha=d_0 a_1 d_1 \dots a_n d_n$. From Lemma \ref{lem-nrra-pos-bnd}, we know that for every $i: 0 \le i \le n$, $Pos^f_\rho[\alpha,2i] \cup Pos^p_{\rho}[\alpha,2i]$ contains only a bounded number of positions.  It follows that only a bounded number of registers are needed to store them in the position $2i$. Therefore, $\Bb$ can simulate $\rho$ as follows: In the position $2i$,
\begin{itemize}
\item  $\Bb$ records in its registers the data values in the positions belonging to $Pos^p_\rho[\alpha,2i]$. 
\item $\Bb$ guesses in its registers the data values in the positions belonging to $Pos^f_\rho[\alpha,2i]$.
\item $\Bb$ records the order for the positions in $Pos^p_\rho[\alpha,2i]$ and $Pos^f_\rho[\alpha,2i]$.
\end{itemize}


We introduce some additional notations.

Let $\alpha=d_0 a_1 d_1 \dots a_n d_n$ be a data path and $i: 0 \le  i \le n$.  The \emph{profile} of the position $2i$ in $\alpha$, denoted by $prof_{\alpha}(2i)$, is defined as a triple $(S,\chi, \sim)$, where
\begin{itemize}
\item $S=\{t \in \Tt_\Aa \mid t_{\alpha}[2i] \neq \bot\}$,
\item $\chi$ is a sequence
\[\begin{array}{l}(b_{-m_1}, T_{-m_1}, b'_{-m_1},s_{-m_1}) \dots (b_{-1}, T_{-1}, b'_{-1},s_{-1})
(b_0, T_0,b'_0, s_{0}) (b_1, T_1,b'_1, s_1) \dots(b_{m_2}, T_{m_2},b'_{m_2},s_{m_2})\end{array}\]
where
\begin{itemize}
\item for every $j: -m_1 \le j \le m_2$, $T_j \subseteq S$ and $T_j \neq \emptyset$,
\item the collection $T_{-m_1},\dots,T_0, \dots,T_{m_2}$ forms a partition of $S$, 
\item for every $t,t' \in S$, if $t \in T_{j_1}$ and $t' \in T_{j_2}$, then $j_1 \le j_2$ iff $t_{\alpha}[2i] \le t'_{\alpha}[2i]$ (in particular, $j_1 = j_2$ iff $t_{\alpha}[2i] = t'_{\alpha}[2i]$),
\item $\cur \in T_0$,
\item for every $j: -m_1 < j \le m_2$, $b_j=a_{(t_{\alpha}[2i])/2}$ for some $t \in T_j$, and $b_{-m_1}=a_{(t_{\alpha}[2i])/2}$ if $t_\alpha [2i]> 0$ for some $t \in T_{-m_1}$, otherwise, $b_{-m_1}=\bot$,
\item for every $j: -m_1 \le j < m_2$, $b'_j=a_{(t_{\alpha}[2i])/2+1}$ for some $t \in T_j$, and $b'_{m_2}=a_{(t_{\alpha}[2i])/2+1}$ if $t_\alpha [2i]< 2n$ for some $t \in T_{m_2}$, otherwise, $b'_{m_2}=\bot$,
\item $s_{m_2}=\bot$, and for every $j: -m_1 \le j < m_2$, if $t'_{\alpha}(2i)=t_{\alpha}(2i)+1$ for some $t \in T_j$ and $t' \in T_{j+1}$, then $s_j = 1$, otherwise, $s_j=0$.
\end{itemize}
\item $\sim$ is an equivalence relation over $S$ defined as follows: Let $t,t'\in S$, then $t \sim t'$ iff $d_{t_{\alpha}[2i]}=d_{t'_{\alpha}[2i]}$.
\end{itemize}

Let $\Sigma_{prof}$ denote the set of all triples $(S,\chi,\sim)$ such that 
\begin{itemize}
\item $S \subseteq \Tt_\Aa$,
\item $\chi$ is a sequence
\[\begin{array}{l}(b_{-m_1}, T_{-m_1}, b'_{-m_1}, s_{-m_1}) \dots (b_{-1}, T_{-1}, b'_{-1},s_{-1})
(b_0, T_0,b'_0,s_{0}) (b_1, T_1,b'_1,s_1) \dots(b_{m_2}, T_{m_2},b'_{m_2},s_{m_2})\end{array}\]
 such that 
 \begin{itemize}
\item for every $j: -m_1 \le j \le m_2$, $T_j \neq \emptyset$, 
\item $\cur \in T_0$, 
\item $T_{-m_1},\dots,T_{m_2}$ is a partition of $S$, 
\item $b_{-m_1} \in \Sigma^\pm \cup \{\bot\}$, and for every $j: -m_1 < j \le m_2$, $b_j \in \Sigma^\pm$, 
\item $b'_{m_2} \in \Sigma^\pm \cup \{\bot\}$, and for every $j: -m_1\le j < m_2$, $ b'_j \in \Sigma^\pm$, 
\item $s_{m_2}=\bot$, and for every $j: -m_1 \le j < m_2$, $s_j \in \{0,1\}$,
\end{itemize}
\item $\sim$ is an equivalence relation over $S$ such that for every $t,t' \in \Tt_\Aa$, if $t,t' \in T_j$ for some $j$, then $t \sim t'$.
\end{itemize}

Note that for $(S,\chi, \sim) \in \Sigma_{prof}$, there may be no data paths  $\alpha$ and a position in $\alpha$ such that the profile of the position in $\alpha$ is $(S,\chi,\sim)$. Nevertheless, we are able to define a consistency condition on the elements from $\Sigma_{prof}$ so that a consistent element from $\Sigma_{prof}$ indeed corresponds to the profile of a position in some data path. Moreover, for two consistent elements from $\Sigma_{prof}$, say $(S_1,\chi_1,\sim_1), (S_2, \chi_2,\sim_2)$, and $a \in \Sigma^\pm$, we are able to define a syntactic successor relation $(S_1,\chi_1, \sim_1)\stackrel{a}{\longrightarrow}(S_2, \chi_2,\sim_2)$, which mimics the changes from $prof_{\alpha}(2i)$ to $prof_{\alpha}(2(i+1))$ by reading a word symbol $a$ in the position $2i+1$ of a data path. 

For $A \subseteq \Sigma^\pm$, a sequence
\[\chi =\begin{array}{l}(b_{-m_1}, T_{-m_1}, b'_{-m_1}, s_{-m_1}) \dots (b_{-1}, T_{-1}, b'_{-1},s_{-1})
 (b_0, T_0,b'_0,s_{0})  (b_1, T_1,b'_1,s_1) \dots(b_{m_2}, T_{m_2},b'_{m_2},s_{m_2})\end{array},\]
and $j: -m_1 \le j \le m_2$, $A$ is said to occur after (resp. before) $T_j$ in $\chi$ if $b_{j'} \in A$ for some $j': j < j' \le m_2$ or $b'_{j'} \in A$ for some $j': j \le j' \le m_2$ (resp. $b'_{j'} \in A$ for some $j': -m_1 \le j' < j$  or $b_{j'} \in A$ for some $j': -m_1 \le j' \le j$).

Let $(S,\chi, \sim) \in \Sigma_{prof}$ and
\[\chi =\begin{array}{l}(b_{-m_1}, T_{-m_1}, b'_{-m_1},s_{-m_1}) \dots (b_{-1}, T_{-1}, b'_{-1},s_{-1})
 (b_0, T_0,b'_0,s_{0}) (b_1, T_1,b'_1,s_1) \dots(b_{m_2}, T_{m_2},b'_{m_2},s_{m_2})\end{array}.\]
Then $(S,\chi,\sim)$ is said to be \emph{consistent} if $\chi$ satisfies the following conditions.
\begin{itemize}
%
%
\item For every $\suc(t) \in \Tt_\Aa$ (resp. $\pred(t) \in \Tt_\Aa$), and every $j: -m_1 \le j < m_2$ (resp. $j: -m_1 < j \le m_2$), $t \in T_j$  iff $\suc(t) \in T_{j+1}$ (resp. $t \in T_j$ iff $\pred(t) \in T_{j-1}$). 
\item For every $\suc(t) \in \Tt_\Aa$ and every $j: -m_1 \le j < m_2$, if $t \in T_j$ and $\suc(t) \in T_{j+1}$, then  $b'_j=b_{j+1}$ and $s_{j}=1$.
\item For every $\pred(t) \in \Tt_\Aa$ and every $j: -m_1 < j \le m_2$, if $t \in T_j$ and $\pred(t) \in T_{j-1}$, then  $b'_{j-1}=b_{j}$ and $s_{j-1}=1$.
\item For every $\suc_A(t) \in \Tt_\Aa$,  if $\suc_A(t) \in T_{j}$ for some $j: -m_1< j \le m_2$, then $b_j \in A$.
\item   For every $\pred_A(t) \in \Tt_\Aa$, if $\pred_A(t) \in T_{j}$ for some $j: -m_1\le j < m_2$, then $b'_{j} \in A$.
\item For every $\suc_A(t) \in \Tt_\Aa$ (resp. $\pred_A(t) \in \Tt_\Aa$), if $\suc_A(t) \in T_{j}$ (resp. $\pred_A(t) \in T_{j}$), then $t \in T_{j'}$ for some $j': j' < j$ (resp. $j': j' > j$).
%
%
\item For every $\suc_A(t) \in \Tt_\Aa$, if $t \in T_{j_1}$ and $\suc_A(t) \in T_{j_2}$ ($j_1 < j_2$), then for every $j_3: j_1 < j_3 < j_2$, $b_{j_3} \nin A$, and for every $j_3: j_1 \le j_3 < j_2-1$, $b'_{j_3} \nin A$; in addition, $b'_{j_2-1} \in A$ implies $s_{j_2-1}=1$.
\item For every $\pred_A(t) \in \Tt_\Aa$, if $\pred_A(t) \in T_{j_1}$ and $t \in T_{j_2}$ ($j_1 < j_2$), then for every $j_3: j_1 < j_3 < j_2$, $b'_{j_3} \nin A$, and for every $j_3: j_1 +1< j_3 \le j_2$, $b_{j_3} \nin A$; in addition, $b_{j_1+1} \in A$ implies $s_{j_1}=1$.
\item For every $t,\suc_A(t) \in \Tt_\Aa$, if $t \in T_j$ for $j: -m_1 \le j \le m_2$,  and $A$ occurs after $T_j$, then $\suc_A(t) \in T_{j'}$ for some $j': j' > j$.
\item For every $t,\pred_A(t) \in \Tt_\Aa$, if $t \in T_j$ for $j: -m_1 \le j \le m_2$,  and $A$ occurs before $T_j$, then $\pred_A(t) \in T_{j'}$ for some $j': j' < j$.
\end{itemize}

\smallskip

\noindent{\bf Claim}. {\it Suppose $(S,\chi,\sim) \in \Sigma_{prof}$. Then $(S,\chi,\sim)$ is consistent iff there is a data path $\alpha$ and a position $2i$ in $\alpha$ such that $prof_{\alpha}(2i)=(S,\chi, \sim)$.}

\begin{proof}

The ``if'' direction is trivial.

The ``only if'' direction: 

Suppose $(S,\chi,\sim)$ is consistent. Let 
\[\chi =\begin{array}{l}(b_{-m_1}, T_{-m_1}, b'_{-m_1},s_{-m_1}) \dots (b_{-1}, T_{-1}, b'_{-1},s_{-1}) 
 (b_0, T_0,b'_0,s_{0}) (b_1, T_1,b'_1,s_1) \dots(b_{m_2}, T_{m_2},b'_{m_2},s_{m_2})\end{array}.\]
For each $T_{j}$, we assign a data value $d_j$, in a way that respects the equivalence relation $\sim$, that is, if $t \in T_j$, $t' \in T_{j'}$, and $t \sim t'$, then $d_j=d_{j'}$. In addition, let $d$ be a data value different from all these $d_j$'s.

For every $j: -m_1 \le j \le m_2$, we define a data path $\alpha_j$ as follows.
\begin{itemize}
\item For $j=-m_1$,
\begin{itemize}
\item if $s_{-m_1} =1$ and $b_{-m_1} \neq \bot$, then let $\alpha_{-m_1}=d b_{-m_1} d_{-m_1}$, 
\item if $s_{-m_1} =1$ and $b_{-m_1} = \bot$, then let $\alpha_{-m_1}=d_{-m_1}$,
\item if $s_{-m_1} =0$ and $b_{-m_1} \neq \bot$, then let $\alpha_{-m_1}= d b_{-m_1} d_{-m_1} b'_{-m_1} d$,
\item if $s_{-m_1} =0$ and $b_{-m_1} = \bot$, then let $\alpha_{-m_1}=d_{-m_1} b'_{-m_1} d$.
\end{itemize}
\item For $j: -m_1 < j < m_2$, 
\begin{itemize}
\item if $s_{j-1} =1$ and $s_j=1$, then let $\alpha_j=d_{j-1} b_{j} d_j$, 
\item if $s_{j-1}=1$ and $s_j=0$, then let $\alpha_j = d_{j-1} b_j d_j b'_j d$,
\item if $s_{j-1}=0$ and $s_j =1$, then let $\alpha_j = d b_j d_j$,
\item if $s_{j-1}=0$ and $s_j =0$, then let $\alpha_j = d b_j d_j b'_j d$.
\end{itemize}
\item For $j=m_2$,
\begin{itemize}
\item if $s_{m_2-1}=1$ and $b'_{m_2}\neq \bot$, then let $\alpha_{m_2}=d_{m_2-1} b_{m_2} d_{m_2} b'_{m_2} d$,
\item if $s_{m_2-1}=1$ and $b'_{m_2} =\bot$, then let $\alpha_{m_2}=d_{m_2-1} b_{m_2} d_{m_2}$,
\item if $s_{m_2-1}=0$ and $b'_{m_2}\neq \bot$, then let $\alpha_{m_2}=d b_{m_2} d_{m_2} b'_{m_2} d$,
\item if $s_{m_2-1}=0$ and $b'_{m_2}= \bot$, then let $\alpha_{m_2}=d b_{m_2} d_{m_2}$.
\end{itemize}
\end{itemize}

Consider the data path $\alpha= \alpha_{-m_1} \cdot \alpha_{-m_1+1} \cdot {\dots} \cdot \alpha_{-1} \cdot \alpha_0 \cdot \alpha_1 \cdot {\dots}\cdot \alpha_{m_2}$.

For every $j: -m_1 \le j \le m_2$, let the position of $\alpha$ corresponding to the data value $d_j$ be $2i_j$.

From the construction of $\alpha$ from $(S,\chi,\sim)$, by an induction on the structure of position terms, we can prove the following result.
\begin{quote} 
\it For every $t \in \Tt_\Aa$ and every $j: -m_1 \le j \le m_2$, $t \in T_j$ iff $t_\alpha[2i_0]=2i_j$. \hfill ($\ast$) 
\end{quote}
Let us take $t=\suc_A(\cur)$ as an example to illustrate the proof. 

Suppose $\suc_A(\cur) \in T_j$, then $b_j \in A$ and for every $j': 0 < j' < j$, $b_{j'} \nin A$, and for every $j': 0 \le j' < j-1$, $b'_{j} \nin A$; in addition, $b'_{j-1} \in A$ implies $s_{j-1}=1$. From this, we deduce that $(\suc_A(\cur))_\alpha[2i_0]=2i_j$, since all the word symbols located after the position $2i_0$ and before the position $2(i_j-1)$ in $\alpha$ do not belong to $A$.

On the other hand, suppose $(\suc_A(\cur))_\alpha[2i_0]=2i_j$, then the word symbol immediately before the position $2i_j$, that is, $b_j$, belongs to $A$, and all the word symbols located after the position $2i_0$ and before the position $2(i_j-1)$ in $\alpha$ do not belong to $A$. From the construction of $\alpha$, it follows that $b_j \in A$ and for every $j': 0 < j' < j$, $b_{j'} \nin A$, and for every $j': 0 \le j' < j-1$, $b'_{j} \nin A$; in addition, $b'_{j-1} \in A$ implies $s_{j-1}=1$ and $b'_{j-1}=b_j$. From this, we conclude that $\suc_A(\cur) \in T_j$.

From the result ($\ast$), we conclude that $prof_\alpha[2i_0]=(S,\chi,\sim)$.
\end{proof}

\smallskip

Let $Prof_{\Aa}$ denote the set of elements of $\Sigma_{prof}$ that are consistent. Suppose $(S,\chi,\sim) \in Prof_\Aa$, 
\[\chi =\begin{array}{l}(b_{-m_1}, T_{-m_1}, b'_{-m_1},s_{-m_1}) \dots (b_{-1}, T_{-1}, b'_{-1},s_{-1})
 (b_0, T_0,b'_0,s_{0}) (b_1, T_1,b'_1,s_1) \dots(b_{m_2}, T_{m_2},b'_{m_2},s_{m_2})\end{array},\]
and $c \in \Cc_\Aa$. Then the satisfaction of $c$ over $(S,\chi,\sim)$, denoted by $(S,\chi,\sim) \models c$, can be defined by interpreting $c$ over $(S,\chi,\sim)$ in a natural way. For instance, if $c=t_1 \sim t_2$, then $(S,\chi,\sim) \models c$ if $t_1,t_2 \in S$ and $t_1 \sim t_2$.

Suppose $a \in \Sigma^\pm$, $(S_1,\chi_1, \sim_1), (S_2,\chi_2,\sim_2) \in Prof_\Aa$, for $i=1,2$,
\[\chi_i =\begin{array}{l}(b_{i,-m_{i,1}}, T_{i,-m_{i,1}}, b'_{i,-m_{i,1}},s_{i,-m_{i,1}}) \dots 
(b_{i,-1}, T_{i,-1}, b'_{i,-1},s_{i,-1}) \\
(b_{i,0}, T_{i,0},b'_{i,0},s_{i,0})
(b_{i,1}, T_{i,1},b'_{i,1},s_{i,1}) \dots(b_{i,m_{i,2}}, T_{i,m_{i,2}},b'_{i,m_{i,2}},s_{i,m_{i,2}})\end{array}.\]
In the following, we will define the concept that $(S_2,\chi_2,\sim_2)$ is a \emph{successor} of $(S_1,\chi_1, \sim_1)$ with respect to $a$, denoted by $(S_2,\chi_2, \sim_2)\xrightarrow{a}(S_1, \chi_1,\sim_1)$.

For every $j: -m_{1,1} \le j \le m_{1,2}$, construct $T'_{1,j} \subseteq \Tt_\Aa$ from $T_{1,j}$ as follows.
\begin{itemize}
\item For every $t \in T_{1,j}$ such that $\suc(\cur) \preceq t$, \ let $t[\suc(\cur) \backslash \cur] \in T'_{1,j}$.
\item For every $t \in T_{1,j}$ and $A \subseteq \Sigma^\pm$ such that $a \in A$ and $\suc_A(\cur) \preceq t$, let $t[\suc_A(\cur)\backslash \cur] \in T'_{1,j}$.
\item For every $t \in T_{1,j}$ such that $t[\cur \backslash \pred(\cur)] \in \Tt_\Aa$, let $t[\cur \backslash \pred(\cur)] \in T'_{1,j}$.
\item For every $t \in T_{1,j}$ and $A \subseteq \Sigma^\pm$ such that $a \in A$ and $t[\cur \backslash \pred_A(\cur)] \in \Tt_\Aa$, let  $t[\cur \backslash \pred_A(\cur)] \in T'_{1,j}$.
\item For every $t \in T_{1,j}$ such that $a \nin A$, $\suc_A(\cur) \preceq t$ or $\pred_A(\cur) \preceq t$, let $t \in T'_{1,j}$.
\item If $s_{1,0}=1$, let $\cur \in T'_{1,1}$.
\end{itemize}

Note that $T'_{1,j}$'s defined above may be empty for some $j: -m_1 \le j \le m_2$.

$(S_1,\chi_1, \sim_1)\stackrel{a}{\longrightarrow}(S_2, \chi_2,\sim_2)$ if the following conditions hold.
\begin{itemize}
\item $b'_{1,0}=a$.
\item Let $j_1j_2 \dots j_\ell: -m_{1,1} \le j_1 < j_2 < \dots < j_\ell \le 0$ be the sequence of the \emph{non-positive} indices such that for every $r: 1 \le r \le \ell$, $T'_{1,j_r} \neq \emptyset$ (and all the other $T'_{1,j}$'s for non-positive $j$'s are empty). Then the sequence 
\[\begin{array}{c} (b_{2,-m_{2,1}}, T_{2,-m_{2,1}}, b'_{2,-m_{2,1}},s_{2,-m_{2,1}}) \dots
(b_{2,-1}, T_{2,-1}, b'_{2,-1},s_{2,-1})\end{array}\]
is equal to the sequence 
$\begin{array}{c} (b_{1,j_1}, T'_{1,j_1}, b'_{1,j_1},s_{1,j_1}) \dots 
(b_{1,j_\ell}, T'_{1,j_\ell}, b'_{1,j_\ell},s_{1,j_\ell}).\end{array}$
\item
There is a partial mapping $f$ from $\{1,\dots,m_{1,2}\}$ to $\{0,\dots,m_{2,2}\}$ such that 
\begin{itemize}
\item for every $j: 1 \le j \le m_{1,2}$, $f(j)$ is undefined iff $T'_{1,j}=\emptyset$,
\item $f$ is increasing, that is, if $j_1 < j_2$ and $f(j_1),f(j_2)$ are defined, then $f(j_1)< f(j_2)$,
\item for every $j: 1 \le j \le m_{1,2}$, if $f(j)$ is defined, then $T'_{1,j} \subseteq T_{2,f(j)}$, $b_{1,j}=b_{2,f(j)}$ and $b'_{1,j}=b'_{2,f(j)}$,
\item if $f(1)$ is defined, then $f(1)=0$ iff $s_{1,0}=1$,
\item for every $j: 1 \le j < m_{1,2}$, if $f(j), f(j+1)$ are both defined, then $s_{1,j}=1$ implies $f(j+1)=f(j)+1$ and $s_{2,f(j)}=1$,
\item for every $j_1,j_2: j_1 < j_2$, if $f(j_1),f(j_2)$ are both defined, then for every $t \in T_{2,f(j_1)}, t' \in T_{2,f(j_2)}$, $t \sim_2 t'$ iff there are $t_1 \in T_{1,j_1}, t'_1 \in T_{1,j_2}$ such that $t_1 \sim_1 t'_1$. 
\end{itemize}
\end{itemize}

Intuitively, $(S_1,\chi_1,\sim_1)$ is rotated one-position to the left to get the profile $(S_2,\chi_2,\sim_2)$. The $T'_{1,j}$'s together with $f$ above define the information that should be inherited during the rotation. 

\smallskip

We are ready to construct the NRAG $\Bb$. 

There are $2|\Tt_\Aa|+1$ registers in $\Bb$, that is, 
\[r_1,\dots, r_{|\Tt_\Aa|}, r_{|\Tt_\Aa|+1}, \dots, r_{2|\Tt_\Aa|}.\]

Over a data path $\alpha=d_0 a_1 d_1 \dots a_n d_n$, $\Bb$ does the following.
\vspace*{-1mm}
\begin{itemize}
\item In each position $2i$ ($0 \le i \le n$), $\Bb$ guesses $\pi_i =(S_i,\chi_i,\sim_i) \in Prof_\Aa$ (where $\pi_i$ is supposed to be $prof_\alpha[2i]$). In addition, 
\begin{itemize}
\item if $i=0$, then $\pi_0=(S_0,\chi_0,\sim_0)$ is an initial profile, that is, for every $t \in \Tt_\Aa$ such that $\pred(\cur) \preceq t$ or $\pred_A(\cur) \preceq t$ for some $A \subseteq \Sigma^\pm$,  $t \nin S_0$,
\item if $i=n$, then $\pi_n=(S_n,\chi_n,\sim_n)$ is a final profile, that is, for every $t \in \Tt_\Aa$ such that $\suc(\cur) \preceq t$ or $\suc_A(\cur) \preceq t$ for some $A\subseteq \Sigma^\pm$, $t \nin S_n$. 
\end{itemize}
\vspace*{-2mm}
\item For every $i: 0 \le i \le n$, if 
\[\chi_i =\begin{array}{l}(b_{i,-m_{i,1}}, T_{i,-m_{i,1}}, b'_{i,-m_{i,1}},s_{i,-m_{i,1}}) \dots \\
(b_{i,-1}, T_{i,-1}, b'_{i,-1},s_{i,-1}) (b_{i,0}, T_{i,0},b'_{i,0},s_{i,0}) \\
(b_{i,1}, T_{i,1},b'_{i,1},s_{i,1}) \dots\\
(b_{i,m_{i,2}}, T_{i,m_{i,2}},b'_{i,m_{i,2}},s_{i,m_{i,2}})\end{array},\]
then after the position $2i$ is visited (that is, the reading head is in $2i+1$), for each $j: -m_{i,1} \le j \le m_{i,2}$, $\Bb$ stores in the register $r_{j+|\Tt_\Aa|}$ the data value corresponding to $T_{i,j}$. In particular, $\Bb$ stores the data value $d_i$ in $r_{|\Tt_\Aa|}$.
\vspace*{-1mm}
\item Over each pair of positions $2i$ and $2(i+1)$ (where $0 \le i < n$), $\Bb$ checks that $\pi_{i} \xrightarrow{a_{i+1}} \pi_{i+1}$. To do this, $\Bb$ copies (by guessing) data values between registers and guesses some data values for a few registers.
\vspace*{-1mm}
\item At the same time, $\Bb$ simulates the run of $\Aa$ as follows.
\begin{itemize}
\item If $\Aa$ makes a transition $(q,a_i,q')$ over $a_i$, then $\Bb$ checks that $b'_{i-1,0}=a_i$ and changes the state from $q$ to $q'$.
\item If $\Aa$ makes a transition $(q,c,q')$ over $d_i$, then $\Bb$ checks that $\pi_i$ satisfies $c$, verifies that $d_i$ is equal to the data value stored in $r_{j+|\Tt_\Aa|}$ for each $j: -m_1 \le j \le m_2$ such that there is $t \in T_j$ satisfying $\cur \sim_i t$ (in particular, $d_i$ should be equal to the data value in $r_{|\Tt_\Aa|}$), and changes the state from $q$ to $q'$.
\item $\Bb$ accepts if $\Aa$ accepts and a final profile is reached.
\end{itemize}
\end{itemize}
From the above construction, we know that in its states, $\Bb$ should record the states of $\Aa$ and the guessed profiles. Therefore, the number of states of $\Bb$ is polynomial over $|Q|$ and exponential over $|\Tt_\Aa|$. \qed
\end{proof}

\smallskip

\noindent {\bf Proposition \ref{prop-nonempt-nrra}}
{\it The nonemptiness of NRRAs and DRRAs is PSPACE-complete.}

\begin{proof}
The upper bound: 

From Theorem \ref{thm-nrra-nrag}, given a NRRA $\Aa=(Q,\delta,I,F)$, an equivalent NRAG $\Bb=(Q',k,\delta',I',F')$ can be constructed such that $\Ll(\Aa)$ is nonempty iff $\Ll(\Bb)$ is nonempty. Moreover,  $\Bb$ satisfies that $|Q'|$ is polynomial over $|Q|$ and exponential over $|\Tt_\Aa|$, and $k$ is polynomial over $|\Tt_\Aa|$. 

From the proof of Proposition \ref{prop-nonempt-nrag}, we know that a NFA $\Bb'$ can be constructed from $\Bb$ such that $\Ll(\Bb')$ is nonempty iff $\Ll(\Bb)$ is nonempty. Since the number of states of $\Bb'$ is polynomial over $|Q'|$ and exponential over $k$, it follows that the number of states of $\Bb'$ is polynomial over $|Q|$ and exponential over $|\Tt_\Aa|$. To decide the nonemptiness of $\Aa$, an accepting run of $\Bb'$ can be guessed nondeterministically in polynomial space. The PSPACE upper bound then follows from Savitch's theorem. 

The lower bound:

The reduction from the membership problem of polynomial space Turing machines to the nonemptiness problem of NRAs or DRAs (\cite{DL09}) can be adapted to a reduction to the nonemptiness problem of NRRAs or DRRAs. 
\end{proof}

\smallskip

\noindent {\bf Proposition \ref{prop-nrra-drra}}
{\it For every NRRA $\Aa$, there is an equivalent DRRA of exponential size.}

\begin{proof}
Let $\Aa=(Q,\delta,I,F)$ be a NRRA. We construct a DRRA $\Aa'=(Q',\delta', q'_0, F')$ as follows:
\begin{itemize}
\item $Q'=Q'_w \cup Q'_d$, where $Q'_w=2^{Q_w}$, $Q'_d=2^{Q_d}$, 
\item $q'_0=I, F'=\{S \in Q'_w \mid S \cap F \neq \emptyset\}$,
\item $\delta'=\delta'_w \cup \delta'_d$ is defined as follows: 
\begin{itemize}
\item $\delta'_w=\{(S,a,S') \mid S \in Q'_w, S' \in Q'_d, S'=\{q' \mid \exists q \in S. (q,a,q') \in \delta_w\}\}$,
\item $\delta'_d$ is defined as follows:\\
For every $S \in Q'_d$, let $C$ denote the set of rigid data constraints occurring in the tuples $(q,c,q') \in \delta_d$ such that $q \in S$. Then $\delta'_d$ contains all tuples $(S,c',S')$ such that there exists $C' \subseteq C$ satisfying that $c'=\bigwedge \limits_{c \in C'} c \wedge \bigwedge \limits_{c \in C \setminus C'} \bar{c}$, and $S'=\{q' \mid \exists q \in S, c \in C'. (q,c,q') \in \delta_d\}$.
\end{itemize}
\end{itemize}
Note that the transitions $(S,c',S')$ may be non-applicable if $c'$ is unsatisfiable.

If $(S,c'_1,S'_1), (S,c'_2,S'_2) \in \delta'_d$ such that $S'_1 \neq S'_2$, then there are $C'_1,C'_2$ such that $C'_1 \neq C'_2$, $c'_1= \bigwedge \limits_{c \in C'_1} c \wedge \bigwedge \limits_{c \in C \setminus C'_1} \bar{c}$ and $c'_2=\bigwedge \limits_{c \in C'_2} c \wedge \bigwedge \limits_{c \in C \setminus C'_2} \bar{c}$. It is easy to observe that if $C'_1 \neq C'_2$, then $c'_1 \wedge c'_2$ is unsatisfiable. Therefore, $\Aa'$ is a DRRA.
\end{proof}

\smallskip

\noindent {\bf Corollary \ref{cor-inc-nrra}}
{\it The language inclusion problem for NRRAs is PSPACE-complete.}

\begin{proof}
The upper bound:

Let $\Aa=(Q_1,\delta_1,I_1,F_1)$ and $\Bb=(Q_2,\delta_2,I_2,F_2)$ be two NRRAs. To decide whether $\Ll(\Aa) \subseteq \Ll(\Bb)$, we use the following procedure.
\begin{itemize}
\item Determinize and complement $\Bb$, let $\Cc$ be the resulting DRRA. 
\item Construct the product of $\Aa$ and $\Bb$, say $\Cc'$, that defines $\Ll(\Aa) \cap \Ll(\Cc)$. Check whether $\Ll(\Cc') \neq \emptyset$. 
\end{itemize}

From the proof of Proposition \ref{prop-nrra-drra}, we know that the size of $\Cc$ is exponential over $|Q_2|$. Thus, the size of $\Cc'$ is polynomial over $|Q_1|$ and exponential over $|Q_2|$. The set of position terms of $\Cc'$ is the union of $\Tt_\Aa$ and $\Tt_\Bb$. 

From the proof of Proposition \ref{prop-nonempt-nrra}, it follows that the nonemptiness of $\Cc'$ can be reduced to that of a NFA of size polynomial over $|Q_1|$, exponential over $|Q_2|$, and exponential over $|\Tt_\Aa|, |\Tt_\Bb|$.

From Savitch's theorem, we conclude that $\Ll(\Aa)\subseteq \Ll(\Bb)$ can be decided in PSPACE.

The lower bound:

The language inclusion of NFAs is already PSPACE-hard.
\end{proof}

\section{Proofs in Section 3.3} 

\noindent{\bf Proposition \ref{prop-2nrra-nrra}}.
{\it For every 2NRRA, there is an equivalent NRRA of exponential size.}

\begin{proof}
The proof is an adaptation of Shepherdson's method \cite{She59} to construct an equivalent NFA from a two-way NFA.

Let $\Aa=(Q,\vdash,\dashv,\delta,I,F)$ be a 2NRRA.  We construct a NRRA $\Aa'=(Q',\delta',I',F')$ as follows.

\begin{itemize}
\item $Q'=Q'_d \cup Q'_w$, where 
\begin{itemize}
\item $Q'_d$ is the set of all tuples $(C,f) \in 2^{\Cc_\Aa} \times (Q \cup \{\bot\})^{Q_d  \cup \{\cdot\}}$ such that $\bigwedge \limits_{c \in C} c$ is satisfiable, $f(\cdot) \in Q_d$, and for every $q \in Q_d$, $f(q) \in Q_w \cup \{\bot\}$, 
\item $Q'_w$ is the set of all tuples  $(a,f) \in (\Sigma^\pm \cup \{\dashv\}) \times (Q \cup \{\bot\})^{Q_w \cup \{\cdot\}}$ such that $f(\cdot) \in Q_w$, and for every $q \in Q_w$, $f(q) \in Q_d \cup \{\bot\}$.
\end{itemize}
\item $I'$ is the set of $(C,f) \in Q'_d$ satisfying the following conditions,
\begin{itemize}
\item $f(\cdot)=q$ such that $(q',\vdash,q,+1) \in \delta_w$ for some $q' \in I$,
\item if $f(q)=q'$, then there exist $q_0,q_1,\dots,q_k \in Q_d$ and $p_1,\dots,p_k \in Q_w$ such that 
\begin{itemize}
\item $q_0=q$, 
\item for every $i: 0 \le i < k$, there is $c_i \in C$ such that $(q_i,c_i,p_{i+1},-1) \in \delta_d$, and for every $i: 1 \le i \le k$, $(p_i,\vdash,q_i,+1)\in \delta_w$,
\item there is $c \in C$ such that $(q_k, c, q',+1) \in \delta_d$.
\end{itemize}
\end{itemize}
\item $\delta'$ are defined as follows. 
\begin{itemize}
\item Let $(a,f)\in Q'_w, (C,f') \in Q'_d$. Then $((a,f),a,(C,f')) \in \delta'_w$ if for every pair $(q,q')$ such that $q \in Q_d$, $q'  \in Q_w$ and $f'(q)=q'$, the following condition holds.
\begin{quote}
There exist $q_0,q_1,\dots,q_k \in Q_d, p_1,\dots,p_k \in Q_w$ such that $q_0=q$, and the following conditions hold,
\begin{enumerate}
\item for every $0 \le i < k$, there exists $c \in C$ such that $(q_i,c,p_{i+1},-1)\in \delta_d$,
\item for every $1 \le i \le k$, $f(p_i)=q_i$,
\item there is $c \in C$ such that $(q_k,c,q',+1)\in \delta$.
\end{enumerate}
\end{quote}
\item Let $(C,f) \in Q'_d, (a,f')\in Q'_w$. Then $((C,f),\bigwedge \limits_{c \in C} c,(a,f'))\in \delta'_d$ iff the following conditions hold.\\
For every pair $(q,q')$ such that $q,q' \in Q$ and $f'(q)=q'$, there exist $q_0,q_1,\dots,q_k \in Q_w$ and $p_1,\dots,p_k \in Q_d$ satisfying that $q_0=q$, and 
\begin{enumerate}
\item for every $0 \le i < k$, $(q_i,a,p_{i+1},-1) \in \delta_w$,
\item for every $1 \le i \le k$, $f(p_i)=q_i$,
\item $(q_k,a,q',+1)\in \delta_w$.
\end{enumerate}
\end{itemize}
\item $F'$ consists of all $(\dashv,f) \in Q'_w$ such that 
\begin{itemize}
\item $f(\cdot)=q \in Q_w$, 
\item for every $q' \in Q_w$, $f(q')=\bot$,
\item there is $(C,f') \in Q'_d$ such that $((C,f'),\bigwedge \limits_{c \in C}c, f) \in \delta'_d$, there are $p_1,\dots,p_k \in Q_d$ and $q_0,\dots,q_k \in Q_w$ satisfying that $q_0=q$, for every $i: 0 \le i < k$, $(q_i,\dashv,p_{i+1},-1)\in \delta_w$, and for every $i: 0 \le i \le k$, $f'(p_i)=q_{i}$, and $q_k \in F$.
\end{itemize}
\end{itemize}

Now we prove the correctness of the construction, that is, for every data path $\alpha=d_0 a_1 d_1\dots a_n d_n$, $\Aa$ accepts $\alpha$ iff $\Aa'$ accepts $\alpha$.

\medskip

\noindent ``\emph{Only if direction}'': 

Suppose $\Aa$ accepts $\alpha$. Then there is an accepting run of $\Aa$ over $\alpha$, say $(q_0,i_0) \theta_0 (q_1,i_1)\theta_1 \dots \theta_{m-1} (q_m,i_m)$, such that 
\begin{itemize}
\item $q_0 \in I$, $q_m \in F$,
\item $i_0=0$, $i_m=2n+2$,
\item for every $j: 0 \le j < m$, if $i_j$ is even, then there is $dir \in \{+1,-1\}$ such that $(q_j,a_{i_j/2},q_{j+1},dir) \in \delta_w$ (where $a_0=\vdash,a_{n+1}=\dashv$), $\theta_j=a_{i_j/2}$, and $i_{j+1}=i_j+dir$, 
\item for every $j: 0 \le j \le m$, if $i_j$ is odd, then there are $c \in \Cc_{rgd}$ and $dir \in \{+1,-1\}$ such that $(q_j,c,q_{j+1},dir) \in \delta_d$, $(\alpha,i_j-1) \models c$, $\theta_j=c$, and $i_{j+1}=i_j + dir$.
\end{itemize}

Without loss of generality, we assume that in the accepting run above, no states are repeated when the reading head moves to the same position, more precisely, the following condition holds. 
\begin{quote}
For every $j_1,j_2: 0 < j_1 < j_2 < m$ such that $i_{j_1}=i_{j_2}$, it holds that $q_{i_{j_1}} \neq q_{i_{j_2}}$. \hfill ($\ast$)
\end{quote}
The above assumption is justified by the fact that if a state is repeated in the same position, then the subrun between the repetitions can be trimmed and the remaining part is still an accepting run.

For each $i: 1 \le i \le 2n+2$, define $f_i$ as follows.
\begin{enumerate}
\item For every $q \in Q$, if there are $j_1,j_2: 0 \le j_1 < j_2 < m$ such that 
\begin{itemize}
\item $q_{j_1}=q$, $q_{j_2}=q'$, 
\item $i_{j_1}=i$, $i_{j_2}=i+1$, and for every $j': j_1 < j' < j_2$, $i_{j'} \le i$,
\end{itemize}
then $f_i(q)=q'$, otherwise $f_i(q)=\bot$. 

Note that the assumption ($\ast$) guarantees that for every $i$, there is at most one pair $(j_1,j_2)$ satisfying the above condition. So $f_i$ is well-defined.

\item $f_i(\cdot)=q$, where $q \in Q$ satisfies that  there exists $j: 0 < j \le m$ such that $i_j=i$, $q=q_{i_j}$, and for every $j': 0 \le j' < j$, $i_{j'} < i$. 
\end{enumerate}
For each $i: 0 \le i \le n$, define $C_i \subseteq \Cc_\Aa$ as the set of $\theta_j$'s such that $i_j=2i+1$. 

Then 
\[(C_0,f_1) (\bigwedge \limits_{c \in C_0} c) (a_1,f_2) a_1 (C_1,f_3) \dots (a_n,f_{2n}) a_n (C_n,f_{2n+1}) (\bigwedge \limits_{c \in C_n} c) (\dashv,f_{2n+2})\]
is an accepting run of $\Aa'$ over $\alpha$.

\medskip

``\emph{If direction}'': 

Suppose 
$(C_0,f_0) c_0 (a_1,f_1) a_1 (C_1,f_2) \dots (a_n,f_{2n-1}) a_n (C_n,f_{2n}) c_n (\dashv,f_{2n+1})$
 is an accepting run of $\Aa'$ over $\alpha$.

Since $f_{2n+1} \in F'$,  there exist $q_0,q_1,\dots,q_k \in Q_w$ and $p_1,\dots,p_k \in Q_d$ such that 
\begin{itemize}
\item $f_{2n+1}(\cdot)=q_0$, 
\item for every $i: 0 \le i < k$, $(q_i,\dashv,p_{i+1},-1)\in \delta_w$, and for every $i: 1 \le i \le k$, $f_{2n}(p_i)=q_i$,
\item $q_k \in F$.
\end{itemize}

From the fact that $f_{2n+1}(\cdot)=q_0 \neq \bot$, we deduce from the definition of $\delta'$ in $\Aa'$ that $f_i(\cdot)\neq \bot$ for every $i: 0 \le i \le 2n$. It follows that for every $i: 1 \le i \le 2n+1$, $f_i(\cdot)=f_{i-1}(f_{i-1}(\cdot))$.

From the fact that $f_0(\cdot) \neq \bot$, there is $p_0 \in I$ such that $(p_0,\vdash, f_0(\cdot), +1)\in \delta_w$.

By induction on $i: 0 \le i \le 2n$, we can show that if $f_i(q)=q'$, then there is a subrun $\rho^i_{q,q'}$ from $(q,i+1)$ to $(q',i+2)$ of $\Aa$ over $\vdash \alpha \dashv$.

Consider the composition of the following subruns,
\[\begin{array}{c} (p_0,0) \vdash (f_0(\cdot),1), \rho^1_{f_0(\cdot),f_0(f_0(\cdot))}, 
\rho^2_{f_1(\cdot),f_1(f_1(\cdot))}, \dots, \rho^{2n}_{f_{2n}(\cdot), f_{2n}(f_{2n}(\cdot))}, 
(q_0,2n+2) \dashv (p_{1},2n+1), \rho^{2n}_{p_1,q_1}, \\
(q_1,2n+2) \dashv (p_2,2n+1), \rho^{2n}_{p_2,q_2}, \dots,
 (q_{k-1},2n+2) \dashv (p_k, 2n+1),\rho^{2n}_{p_k,q_k}.\end{array}\]
Let $\rho$ denote this composition. Then $\rho$ is an accepting run of $\Aa$ over $\alpha$.
\end{proof}

\section{Proofs in Section 4}

\noindent {\bf Proposition \ref{prop-complexity-2rrdpq}}
{\it The evaluation for 2RRDPQs is PSPACE-complete, and NLOGSPACE-complete in data complexity.
}

\begin{proof}

The upper bound: 

We use the idea to prove the PSPACE upper bound for 2RDPQs in \cite{LV12a}.

Let $\dg=(V,E,\eta)$ be a data graph,$\xi=(x,L,y)$ be a 2RRDPQ, and $(v_1,v_2) \in V \times V$. Suppose $L$ is given by a NRRA $\Aa=(Q,\delta,I,F)$ over $\Sigma^\pm$.

Let $D$ be the set of data values occurring in $\dg$. Then $\dg$ plus $(v_1,v_2)$ can be seen as a NFA $\Aa_{\dg,(v_1,v_2)}=(Q', \delta',I',F')$ with initial state $(v_1)_s$ and final state $(v_2)_t$ over the alphabet $\Sigma^\pm \cup D$ as follows.
\begin{itemize}
\item $Q'=\{v_s,v_t \mid v \in V\}$,
\item $\delta'=\{(v_t, a, v'_s),(v'_t, a^-, v_s) \mid (v,a,v')\in E\} \cup \{(v_s,d,v_t) \mid v \in V, \eta(v)=d\}$,
\item $I'=\{v_1\}$, $F'=\{v_2\}$.
\end{itemize}

From the proof of Theorem \ref{thm-nrra-nrag}, we know that from $\Aa$, an equivalent NRAG $\Bb=(Q',k,\delta',I',F')$ can be constructed such that $|Q'|$ is polynomial over $|Q|$ and exponential over $|\Tt_\Aa|$, and $k$ is polynomial over $|\Tt_\Aa|$.

When restricted to the data paths where all data values are from $D$, the NRAG $\Bb$ can be seen as a NFA $\Bb'$ over the alphabet $\Sigma^\pm \cup D$ with the state space $Q' \times D^{[k]}$. It follows that the size of the state space of $\Bb'$ is exponential over the size of $\Aa$ and polynomial over the size of $D$. 

To decide whether $(v_1,v_2) \in \xi(\dg)$, it is sufficient to check whether $\Ll(\Aa_{\dg,(v_1,v_2)} \cap \Bb') \neq \emptyset$. Since an accepting run of $\Aa_{\dg,(v_1,v_2)} \cap \Bb'$ can be guessed in polynomial space, from Savitch's theorem, we conclude that the evaluation problem of NRRAs is in PSPACE.

If the size of $\Aa$ is bounded by a constant, then an accepting run of $\Aa_{\dg,(v_1,v_2)} \cap \Bb'$ can be guessed in logarithmic space, it follows that the upper bound of the data complexity of the evaluation problem of NRRAs is NLOGSPACE.

The PSPACE lower bound is obtained by an easy reduction from the nonemptiness of NRRA. The NLOGSPACE lower bound of data complexity is from that of RPQs.
\end{proof}

\smallskip

{\bf Theorem \ref{thm-rdpq2rrdpq}}.
{\it Let $k \ge 1$, $\xi=(x,L,y)$ be a 2RDPQ over the alphabet $\Sigma$ such that $L$ is given by a NRA or REM containing at most $k$-registers. Then a 2RRDPQ $\xi'=(x,L',y)$ over the alphabet $\Sigma^\pm \cup \{A_i, A_i^-\mid 1 \le i \le k\}$ can be constructed in polynomial time such that for every data graph $\dg=(V,E,\eta)$, $\xi(\dg)=\xi'(\dg_{dn,k})$.}


\begin{proof}
Let $(x,L,y)$ be a 2RDPQ. We first consider the situation that $L$ is given by a NRA $\Aa=(Q,k,\delta,I,F)$ over the alphabet $\Sigma^\pm$.

In the following, we will construct a NRRA $\Aa'=(Q',\delta',I',F')$ over $\Sigma^\pm \cup \{A_i, A_i^- \mid 1 \le i \le k\}$ so that $\xi'=(x,\Ll(\Aa'),y)$ satisfies that $\xi(\dg)=\xi'(\dg_{dn,k})$.

The intuition of $\Aa'$ is to simulate the run of $\Aa$, by using the following tricks.
\begin{quote}
Every time a data value $d$ is stored into the $i$-th register in $\Aa$, the sequence $d A_i d A^-_i d$ is read by $\Aa'$. Later on, we can refer to the data values stored in the $i$-th register by using the position terms $\pred_{A^-_i}$. 
\end{quote}

We formally define $\Aa'=(Q',\delta',I',F')$ as follows.
\begin{itemize}
\item $Q' = Q'_w \cup Q'_d$ such that 
\begin{itemize}
\item $Q'_w$ is the union of $Q_w$ and $\delta_d \times \{A_i,A^-_i \mid 1 \le i \le k\}$,
\item $Q'_d$ is the union of $Q_d$ and $\delta_d \times \{\$_i,\#_i \mid 1 \le i \le k\}$.
\end{itemize}
\item $I'=I, F'=F$.
\item $\delta'=\delta'_w \cup \delta'_d$ is defined as follows.
\begin{itemize}
\item $ \delta_w \subseteq \delta'_w$.
\item For every transition $(q,c,q',X) \in \delta_d$, let  $c' \in \Cc_{rgd}[\Sigma^\pm]$ be obtained from $c$ by replacing every $r_j$ ($1 \le j \le k$) with $\pred_{A^-_j}$ and $r_0$ with $\cur$. If $X=\emptyset$, then $(q,c',q') \in \delta'_d$, otherwise, let $X=\{r_{i_1},\dots,r_{i_\ell}\}$, then $\delta'$ includes the following transitions, 
\[\begin{array}{c}q \xrightarrow{c'} ((q,c,q',X),A_{i_1}) \xrightarrow {A_{i_1}} ((q,c,q',X),\$_1) 
\xrightarrow{true} ((q,c,q',X),A^-_{i_1}) \\ 
\xrightarrow{A^-_{i_1}} ((q,c,q',X), \#_1) 
\xrightarrow{true} ((q,c,q',X),A_{i_2}) \xrightarrow{A_{i_2}} ((q,c,q',X),\$_2) 
\dots \\
\xrightarrow{true} ((q,c,q',X), A_{i_\ell}) \xrightarrow {A_{i_\ell}} ((q,c,q',X), \$_\ell) 
\xrightarrow{true} ((q,c,q',X), A^-_{i_\ell}) \\
\xrightarrow{A^-_{i_\ell}} ((q,c,q',X),\#_\ell) 
\xrightarrow{true} q'
\end{array}\]
\end{itemize}
\end{itemize}

If $L$ is given by a REM $e$ over the alphabet $\Sigma^\pm$, we construct a RREM $e'$ such that $\xi'=(x,L(e'),y)$ over the alphabet $\Sigma^\pm \cup \{A_i,A^-_i \mid 1 \le i \le k\}$ satisfies that $\xi(\dg)=\xi'(\dg_{dn,k})$.

From a REM $e$, we construct a RREM $tr(e)$ by an induction on the structure of REMs. The nontrivial cases are $e=\downarrow_X e_1$ and $e=e_1[c]$. For $e=\downarrow_X e_1$, suppose $X=\{r_{i_1},\dots,r_{i_\ell}\}$, then $tr(e)=A_{i_1} A^-_{i_1}\dots A_{i_\ell} A^-_{i_\ell} tr(e_1)$. For $e=e_1[c]$, let $c' \in C_{rgd}$ be obtained from $c$ by replacing $r_{j}$ with $\pred_{A^-_j}$ and $r_0$ with $\cur$, then $tr(e)=tr(e_1)\cdot [c']$.
\end{proof}

\section{Proofs in Section 5}

\noindent {\bf Proposition \ref{prop-eval-c2rrdpq}}.
{\it The evaluation of C2RRDPQs is PSPACE-complete, and NLOGSPACE-complete in data complexity.}

\begin{proof}
The PSPACE lower bound follows from that of 2RRDPQs. The NLOGSPACE lower bound follows from that of RPQs.

The upper bound: 

Let $\xi: = Ans(\bar{z}) \leftarrow \bigwedge \limits_{1 \le i \le l} (y_{2i-1}, L_i, y_{2i})$ be a C2RRDPQ, $\dg=(V,E,\eta)$ a data graph, and $\overline{v}$ is a tuple of nodes of the same arity as $\overline{z}$. Suppose for every $i: 1 \le i \le l$, $L_i$ is given by a NRRA $\Aa_i=(Q_i,\delta_i,I_i,F_i)$ over the alphabet $\Sigma^\pm$.

From the proof of Theorem \ref{thm-nrra-nrag}, we know that from each $\Aa_i$, an equivalent NRAG $\Bb_i=(Q'_i,k_i,\delta'_i,I'_i,F'_i)$ can be constructed such that $|Q'_i|$ is polynomial over $|Q_i|$ and exponential over $|\Tt_{\Aa_i}|$, and $k$ is polynomial over $|\Tt_{\Aa_i}|$.

When restricted to the data paths where all data values are from $D$, the NRAG $\Bb_i$ can be seen as a NFA $\Bb'_i$ over the alphabet $\Sigma^\pm \cup D$ with the state space $Q'_i \times D^{[k_i]}$. It follows that the size of the state space of $\Bb'_i$ is exponential over the size of $\Aa_i$ and polynomial over the size of $D$. 

To check wether $\bar{v} \in \xi(\dg)$, an assignment $\nu$ of nodes in $V$ to $\{y_1,\dots,y_{2l}\}$ is first guessed such that $\nu(\overline{z})=\overline{v}$. 

Similarly to the proof of Proposition \ref{prop-complexity-2rrdpq}, for every pair $(\nu(y_{2i-1}),\nu(y_{2i}))$,  the data graph $\dg$ together with $(\nu(y_{2i-1}),\nu(y_{2i}))$ can be seen as a NFA $\Aa_{\dg,i}$ over the alphabet $\Sigma^\pm \cup D$ with the initial state $(\nu(y_{2i-1}))_s$ and the final state $(\nu(y_{2i}))_t$.

Then for every $i: 1 \le i \le l$, an accepting run of $\Aa_{\dg,i} \cap \Bb'_i$ can be guessed in polynomial space. To check whether $\overline{v} \in \xi(\dg)$, the accepting runs of the NFAs $\Aa_{\dg,i} \cap \Bb'_i$ can be guessed one by one. From Savitch's theorem, we deduce that the nonemptiness of C2RRDPQs is in PSPACE.

Similarly, if the size of $\xi$ is bounded by a constant, then the assignment $\nu$ and the accepting runs of $\Aa_{\dg,i} \cap \Bb'_i$ can be guessed in logarithmic space. Therefore, the evaluation problem of C2RRDPQs has the NLOGSPACE data complexity. 
\end{proof}

\noindent {\bf Lemma \ref{lem-pt-pt}}.
{\it Suppose $\pi'=v_0 a_1 v_1 \dots a_\ell v_\ell$ is a semipath in $\dg$ such that $urv_{\overline{\pi}}(\pi')=\pi'_0 \# \dots \# \pi'_r$ and $trc(\pi')=p_0 b_1 p_1 \dots b_{\ell+2r} p_{\ell+2r}$ (where $b_1,\dots,b_{\ell+2r} \in \Sigma^\pm \cup \{\#\}$). Then for every $i: 0 \le i \le \ell+2r$, there exists a function $pos_i \in (\Tt_p[\Sigma_{\xi_1}] \cup \{\bot\})^{\Tt_\Aa}$ such that for every $t \in \Tt_\Aa$,  $pos_i(t) =\bot$ iff $t^{adj}_{urv_{\overline{\pi}}(\pi')}[2i]=\bot$; moreover, if $pos_i(t) \neq \bot$ and $t^{adj}_{urv_{\overline{\pi}}(\pi')}[2i]=2i'$, then $(pos_i(t))_{\alpha_\dg}[p_i]=p_{i'}$.
}

\begin{proof}

Let $\pi'=v_0 a_1 v_1 \dots a_\ell v_\ell$ be a semipath in $\dg$, $unr_{\overline{\pi}}(\pi')=\pi'_0 \# \dots \# \pi'_r$ such that for every $s: 0 \le s \le r$, all the edges on $\pi'_s$ belonging to $\pi_{j_s}$ for some $j_s: 1 \le j_s \le l_1$, and $trc(\pi')=p_0 b_1 p_1 \dots b_{\ell+2r} p_{\ell+2r}$.

We prove the lemma by an induction on the structure of position terms.

Induction base: For every $i: 0 \le i \le \ell+2r$, $pos_i(\cur)=\cur$.

Induction step:

Let us first consider the case $t=\suc(t_1)$.

Let $i: 0 \le i \le \ell+2r$.

If $t^{adj}_{urv_{\overline{\pi}}(\pi')}[2i]=\bot$, then let $pos_i(t)=\bot$. Otherwise, $t^{adj}_{urv_{\overline{\pi}}(\pi')}[2i]=2i'$ for some $i'$. From $t=\suc(t_1)$, it follows that $(t_1)^{adj}_{urv_{\overline{\pi}}(\pi')}[2i]=2i''$ for some $i''$ such that $2i'=2i''+2$ or $2i'=2i''+4$.

According to the induction hypothesis, there exists $pos_i(t_1) \in \Tt_p[\Sigma_{\xi_1}]$ such that $(pos_i(t_1))_{\alpha_\dg}[p_{i}]=p_{i''}$.

\begin{itemize}
\item If $2i'=2i''+2$, then let $pos_i(t)=pos_i(\suc(t_1))=\suc(pos_{i}(t_1))$ if $p_{i''} <p_{i'}$, otherwise, let $pos_i(t)=pos_i(\suc(t_1))=\pred(pos_i(t_1))$.
\item If $2i'=2i''+4$, then there are $j_1,j_2: 1 \le j_1, j_2 \le l_1$ such that one of the following conditions holds,
\begin{enumerate}
\item $p_{i''}$ is the position immediately before $\$_{2j_1}$, $p_{i'}$ is the third position before $\$_{2j_2}$  in $\alpha_\dg$, and $j_1 \neq j_2$,
\item $p_{i''}$ is the position immediately before $\$_{2j_1}$, $p_{i'}$ is the third position after $\$_{2j_2-1}$,
\item $p_{i''}$ is the position immediately after $\$_{2j_1-1}$, $p_{i'}$ is the third position before $\$_{2j_2}$,
\item $p_{i''}$ is the position immediately after $\$_{2j_1-1}$, $p_{i'}$ is the third position after $\$_{2j_2-1}$, and $j_1 \neq j_2$.
\end{enumerate}

We illustrate the argument by considering the second situation above. The arguments for the other three situations are similar.

\begin{itemize}
\item if $j_1 < j_2$, then let $pos_i(\suc(t_1))=\suc(\suc_{\$_{2j_2-1}}(pos_i(t_1)))$,
\item if $j_1 > j_2$, then let $pos_i(\suc(\cur))=\suc^2(\pred_{\$_{2j_2-1}}(pos_i(t_1)))$.
\end{itemize}
\end{itemize}

The case $t=\pred(t_1)$ can be discussed similarly as $t=\suc(t_1)$.

Now consider the case $t=\suc_A(t_1)$.

Let $i: 0 \le i \le \ell+2r$.

If $t^{adj}_{urv_{\overline{\pi}}(\pi')}[2i]=\bot$, let $pos_i(t)=\bot$. Otherwise, let $t^{adj}_{urv_{\overline{\pi}}(\pi')}[2i]=2i'$. From $t=\suc_A(t_1)$, we know that $(t_1)^{adj}_{urv_{\overline{\pi}}(\pi')}[2i]=2i''$ for some $i''$ such that $2i''< 2i'$.

From the induction hypothesis, $(pos_i(t_1))_{\alpha_\dg}(p_{i})=p_{i''}$.

If there are no $\#$ symbols in the subpath of $urv_{\overline{\pi}}(\pi')$ from the position $2i''$ to $2i'$, then the position $2i''$ and $2i'$ both belong to $\pi'_s$ for some $s: 0 \le s \le r$. It follows that $p_{i''}$ and $p_{i'}$ are two positions between $\$_{2j_s-1}$ and $\$_{2j_s}$ in $\alpha_\dg$. Define $pos_i(t)$ as follows.
\begin{itemize}
\item If $p_{i'} < p_{i''}$, let $pos_i(\suc_A(t_1))=\pred_{A \times \{j_s\}}(pos_i(t_1))$.
\item If $p_{i''}< p_{i'}$, let $pos_i(\suc_A(t_1))=\suc_{A \times \{j_s\}}(pos_i(t_1))$.
\end{itemize}

Otherwise (that is, there are $\#$ symbols from $2i''$ to $2i'$), let $2i'''$ be the position before the position $2i'$ on $urv_{\overline{\pi}}(\pi')$ such that $2i'''$ is a position immediately after $\#$ and $2i'''$ is the last position before $2i'$ satisfying this property. Let $s: 0 \le s \le r$ such that $2i'''$ and $2i'$ are two positions belonging to $\pi'_s$. Then $p_{2i'''}$ is the position immediately after $\$_{2j_s-1}$ or the position immediately before $\$_{2j_s}$ in $\alpha_\dg$. We illustrate the argument by considering the situation that $p_{2i'''}$ is the position immediately after $\$_{2j_s-1}$. The discussion for the latter situation is similar. Define $pos_i(t)$ as follows.
\begin{itemize}
\item If $p_{i'} < p_{i''}$, let  $pos_i(\suc_A(t_1))=\suc_{A \times \{j_s\}}(\pred_{\$_{2j_s-1}}(pos_i(t_1)))$.
\item If $p_{i''}< p_{i'}$, let 
$pos_i(\suc_A(t_1))=\suc_{A \times \{j_s\}}(\suc_{\$_{2j_s-1}}(pos_i(t_1)))$.
\end{itemize}
 
The case $t=\pred_A(t_1)$ can be discussed similarly to $t=\suc_A(t_1)$.

In summary, for every $t=op(t_1) \in \Tt_p[\Sigma^\pm]$ such that $pos_i(t) \neq \bot$ (where $op=\suc,\pred,\suc_A,\pred_A$), there is $t_{op} \in \Tt_p[\Sigma_{\xi_1}]$ such that $pos_i(t)=t_{op}[\cur \backslash pos_i(t_1)]$.
\end{proof}

\noindent {\bf Theorem \ref{thm-rrdpq-auto}}.
{\it Let $\dg$ be a $\nu$-canonical data graph for $\xi_1$, $\xi$ be a 2RRDPQ. Then a 2NRRA $\Aa_\xi$ can be constructed from $\xi$ and $\xi_1$ such that $\xi(\dg)$ is nonempty iff $\Aa_{\xi}$ accepts $\vdash \alpha_\dg \dashv$.}

\begin{proof}

Let $\pi'$ be a path in $\dg$, the $\overline{\pi}$-unraveling of $\pi$ be $\pi'_0 \dots \pi'_r$, where for every $s: 0 \le s \le r$, all the edges on $\pi'_s=v_{i_s} a_{i_s+1} v_{i_s+1} \dots v_{i_{s+1}}$ belong to $\pi_{j_s}$.


Our goal is to construct a 2NRRA $\Bb$ to simulate the runs of $\Aa'$ over $\eta(urv_{\overline{\pi}}(\pi'))$.

Similarly to the transformation from NRRAs to NRAGs in Theorem \ref{thm-nrra-nrag}, the 2NRRA $\Bb$ goes through $trc(\pi')$ in $\alpha_\dg$ and guesses the profile of the current position of $\eta(urv_{\overline{\pi}}(\pi'))$, in order to simulate $\Aa'$ over $\eta(urv_{\overline{\pi}}(\pi'))$. The difference is that instead of storing and guessing the data values, $\Bb$ records and guesses a position term from $\Tt_p[\Sigma_{\xi_1}]$ (interpreted over $\alpha_\dg$) for each position term occurring in the profile of the current position in $\eta(urv_{\overline{\pi}}(\pi'))$. The intricacy of the construction is how to guarantee the consistency of the guessed position terms $\Tt_p[\Sigma_{\xi_1}]$ and how to update them during the simulation.

A \emph{locating profile} $loc$ of $\Aa'$ over $\alpha_\dg$,  is defined as a pair $((S,\chi,\sim),pos)$, where $(S,\chi,\sim) \in Prof_{\Aa}$ (cf. proof of Theorem \ref{thm-nrra-nrag}), 
\[\chi=\begin{array}{l}(b_{-m_1}, T_{-m_1}, b'_{-m_1},s_{-m_1}) \dots (b_{-1}, T_{-1}, b'_{-1},s_{-1}) \\
(b_0, T_0,b'_0, s_{0}) (b_1, T_1,b'_1, s_1) \dots(b_{m_2}, T_{m_2},b'_{m_2},s_{m_2})\end{array}\]
and $pos: (\Tt_p[\Sigma_{\xi_1}] \cup \{\bot\})^{\Tt_\Aa}$ such that
\begin{itemize}
\item $pos(t)=\bot$ for every $t \in \Tt_\Aa \setminus S$,
\item $pos(\cur)=\cur$,
\item for every $t,t' \in S$ such that $t \preceq t'$, we have $pos(t) \preceq pos(t')$,
\item for every $t_1,t_2 \in S$ such that there is $j: -m_1 \le j \le m_2$ satisfying that $t_1,t_2 \in T_j$, if $op(t_1),op(t_2) \in S$ for $op\in \{\suc,\pred,\suc_A,\pred_A \mid A \subseteq \Sigma^\pm\}$, then there is $t_{op} \in \Tt_p[\Sigma_{\xi_1}]$ of the form as those in the proof of Lemma \ref{lem-pt-pt} (e.g. $t_{suc}=\suc(\suc_{\$_{2j_2-1}}(\cur))$) such that $pos(op(t_j)) = t_{op}[\cur \backslash pos(t_j)]$ for $j=1,2$.
\end{itemize}

Let $\Sigma_{loc}$ denote the set of locating profiles.

Similar to the construction of NRAGs from NRRAs, we define two successor relations between locating profiles.

Let $((S_1,\chi_1,\sim_1),pos_1), ((S_2,\chi_2,\sim_2),pos_2) \in \Sigma_{loc}$, $a \in \Sigma^\pm$, $j: 1 \le j \le l_1$,  $1 \le k_1,k_2 \le 2 l_1$, and $dir\in \{+1,-1\}$. In the following, we will define two relations $((S_1,\chi_1,\sim_1),pos_1) \xrightarrow{((a,j),dir)} ((S_2,\chi_2,\sim_2),pos_2)$ and $((S_1,\chi_1,\sim_1),pos_1) \xrightarrow{(a,\$_{k_1},\$_{k_2})} ((S_2,\chi_2,\sim_2),pos_2)$. The latter relation corresponds to the situation that the run of $\Aa'$ is jumping over $\#$ on $urv_{\overline{\pi}}(\pi')$, and the former relation corresponds to the situation that the run of $\Aa'$ is not.

At first, $((S_1,\chi_1,\sim_1),pos_1) \xrightarrow{((a,j)dir)} ((S_2,\chi_2,\sim_2),pos_2)$ if the following conditions hold.
\begin{itemize}
\item $(S_1,\chi_1,\sim_1) \xrightarrow{a} (S_2,\chi_2,\sim_2)$.
\item If $\suc(\cur) \in S_1$, then $pos_1(\suc(\cur))=\suc(\cur)$ if $dir=+1$, and $pos_1(\suc(\cur))=\pred(\cur)$ otherwise.
\item If $a \in A$ and $\suc_A(\cur) \in S_1$, then $pos_1(\suc_A(\cur))=\suc_{A \times \{j\}}(\cur)$ if $dir=+1$, and  $pos_1(\suc_A(\cur))=\pred_{A \times \{j\}}(\cur)$ otherwise.
\item For every $t \in S_1$ such that $\suc(\cur) \preceq t$, if $dir=+1$, then 
$pos_2(t[\suc(\cur)\backslash \cur])=pos_1(t)[\suc(\cur) \backslash \cur],$
otherwise,  $pos_2(t[\suc(\cur)\backslash \cur])=pos_1(t)[\pred(\cur) \backslash \cur]$.
\item For every $t \in S_1$ such that $\suc_A(\cur) \preceq t$ and $a \in A$, if $dir=+1$, then
\[pos_2(t[\suc_A(\cur) \backslash \cur])=pos_1(t)[\suc_{A \times \{j\}}(\cur) \backslash \cur],\]
otherwise, 
$pos_2(t[\suc_A(\cur) \backslash \cur])=pos_1(t)[\pred_{A \times \{j\}}(\cur) \backslash \cur]$.
\item For every $t \in S_2$ such that $\pred(\cur) \preceq t$, $pos_2(t)=pos_1(t[\pred(\cur)\backslash \cur])$.
\item For every $t \in S_2$ such that $\pred_A(\cur) \preceq t$ and $a \in A$, $pos_2(t)=pos_1(t[\pred_A(\cur)\backslash \cur])$.
\item For every $t \in S_1$ such that $\suc_A(\cur) \preceq t$ and $a \nin A$, $pos_2(t)=pos_1(t)$.
\item For every $t \in S_1$ such that $\pred_A(\cur) \preceq t$ and $a \nin A$, $pos_2(t)=pos_1(t)$.
\end{itemize}

In the following, we will define 
$((S_1,\chi_1,\sim_1),pos_1) \xrightarrow{(a,\$_{k_1},\$_{k_2})} ((S_2,\chi_2,\sim_2),pos_2)$
for $k_1, k_2$ satisfying that there are $k'_1,k'_2: 1 \le k'_1, k'_2 \le l_1$ such that one of the following conditions hold.
\begin{enumerate}
\item $k_1=2k'_1$, $k_2=2k'_2$, and $k'_1 \neq  k'_2$,
\item or $k_1 = 2k'_1$, $k_2=2k'_2-1$,
\item or $k_1 = 2k'_1-1$, $k_2=2k'_2$,
\item $k_1=2k'_1-1$, $k_2=2k'_2-1$, and $k'_1 \neq k'_2$.
\end{enumerate}

We will illustrate the definition for the first case above, the other three cases can be discussed in the same way.
Suppose $k_1=2k'_1$, $k_2=2k'_2$, and $k'_1 \neq  k'_2$ for some $k'_1,k'_2$. Then $((S_1,\chi_1,\sim_1),pos_1) \xrightarrow{(a,\$_{2k'_1},\$_{2k'_2})} ((S_2,\chi_2,\sim_2),pos_2)$
if the following conditions hold.
\begin{itemize}
\item $(S_1,\chi_1,\sim_1) \xrightarrow{a} (S_2,\chi_2,\sim_2)$.
\item If $\suc(\cur) \in S_1$, then 
$pos_1(\suc(\cur))=\pred^2(\suc_{\$_{k_2}}(\cur))$
if $k'_1 < k'_2$, and 
$pos_1(\suc(\cur))=\pred(\pred_{\$_{k_2}}(\cur))$
otherwise.
\item If $a \in A$ and $\suc_A(\cur) \in S_1$, then 
$pos_1(\suc_A(\cur))= \pred_{A \times \{k'_2\}}(\suc_{\$_{k_2}}(\cur))$
if $k'_1 < k'_2$, and  
$pos_1(\suc_A(\cur))=\pred_{A \times [k'_2]}(\pred_{\$_{k_2}}(\cur))$
otherwise.
\item For every $t \in S_1$ such that $\suc(\cur) \preceq t$, if $k'_1 < k'_2$, then 
\[\begin{array}{c}pos_2(t[\suc(\cur)\backslash \cur])=
pos_1(t)[\pred^2(\suc_{\$_{k_2}}(\cur)) \backslash \cur]\end{array},\]
otherwise,  
\[\begin{array}{c}pos_2(t[\suc(\cur)\backslash \cur])=
pos_1(t)[\pred(\pred_{\$_{k_2}}(\cur)) \backslash \cur]\end{array}.\]
\item For every $t \in S_1$ such that $\suc_A(\cur) \preceq t$ and $a \in A$, if $k'_1 < k'_2$, then 
\[\begin{array}{c}pos_2(t[\suc_A(\cur) \backslash \cur])=
pos_1(t)[\pred_{A \times \{k'_2\}}(\suc_{\$_{k_2}}(\cur)) \backslash \cur]\end{array},\]
otherwise, 
\[\begin{array}{c}pos_2(t[\suc_A(\cur) \backslash \cur])=
pos_1(t)[\pred_{A \times [k'_2]}(\pred_{\$_{k_2}}(\cur))\backslash \cur]\end{array}.\]
\item For every $t \in S_2$ such that $\pred(\cur) \preceq t$, $pos_2(t)=pos_1(t[\pred(\cur)\backslash \cur])$.
\item For every $t \in S_2$ such that $\pred_A(\cur) \preceq t$ and $a \in A$, $pos_2(t)=pos_1(t[\pred_A(\cur)\backslash \cur])$.
\item For every $t \in S_1$ such that $\suc_A(\cur) \preceq t$ and $a \nin A$, $pos_2(t)=pos_1(t)$.
\item For every $t \in S_1$ such that $\pred_A(\cur) \preceq t$ and $a \nin A$, $pos_2(t)=pos_1(t)$.
\end{itemize}

We are ready to construct the 2NRRA $\Bb$. 

Suppose $\pi'=v_0 a_1 v_1 \dots a_\ell v_\ell$ is a semipath in $\alpha_\dg$, $urv_{\overline{\pi}}(\pi')=\pi'_0 \# \dots \# \pi'_r$, for every $s: 0 \le s \le r$, all the edges on $\pi'_s$ belong to $\pi_{j_s}$ ($1 \le j_s \le l_1$), and $trc(\pi')=d_0 a'_1 p_1 \dots a'_{\ell+2r} p_{\ell+2r}$ (where for every $j: 1 \le j \le \ell+2r$, $a'_j \in \Sigma^\pm \cup \{\#\}$). Then $\Bb$ does the following.
\begin{itemize}
\item In each position $p_i$ ($0 \le i \le \ell+2r$) of $\alpha_\dg$, $\Bb$ guesses a locating profile $loc_i=((S_i,\chi_i,\sim_i),pos_i) \in \Sigma_{loc}$ with 
\[\chi_i =\begin{array}{l}(b_{i,-m_{i,1}}, T_{i,-m_{i,1}}, b'_{i,-m_{i,1}},s_{i,-m_{i,1}}) \dots 
(b_{i,-1}, T_{i,-1}, b'_{i,-1},s_{i,-1}) \\
(b_{i,0}, T_{i,0},b'_{i,0},s_{i,0}) (b_{i,1}, T_{i,1},b'_{i,1},s_{i,1}) \dots
(b_{i,m_{i,2}}, T_{i,m_{i,2}},b'_{i,m_{i,2}},s_{i,m_{i,2}})\end{array}.\]
In addition, 
\begin{itemize}
\item if $i=0$, then $loc_0=(S_0,\chi_0,\sim_0)$ is an initial locating profile, that is, for every $t \in \Tt_\Aa$ such that $\pred(\cur) \preceq t$ or $\pred_A(\cur) \preceq t$ for some $A \subseteq \Sigma^\pm$,  $t \nin S_0$,
\item if $i=\ell+2r$, then $loc_{\ell+2r}=(S_{\ell+2r},\chi_{\ell+2r},\sim_{\ell+2r})$ is a final profile, that is, for every $t \in \Tt_\Aa$ such that $\suc(\cur) \preceq t$ or $\suc_A(\cur) \preceq t$ for some $A \subseteq \Sigma^\pm$, $t \nin S_{\ell+2r}$. 
\end{itemize}
\item Over each pair of positions $p_{i}$ and $p_{i+1}$ (where $0 \le i < \ell+2r$) of $\alpha_{\dg}$, 
\begin{itemize}
\item if $i_s + 2s \le 2i < i_{s+1} + 2s$ for some $s: 0 \le s \le r$, then $\Bb$ checks that $loc_i \xrightarrow{(a'_{i+1},j_s),dir} loc_{i+1}$, where $dir=+1$ if $p_{i+1}=p_i+2$, and $dir=-1$ otherwise,
\item if $2i=i_s+2(s-1)$ for some $s: 1 \le s \le r$ (that is, $2i$ is the position immediately before $\#$ in $urv_{\overline{\pi}}(\pi')$), then $\Bb$ jumps from $p_{i}$ to $p_{i+1}$, then to $p_{i+2}$, and checks that $loc_i \xrightarrow{(a'_{i+2},\$_k,\$_{k'})} loc_{i+2}$, where 
\begin{quote}
$k=2j_{s-1}$ if $p_{\alpha_\dg}(\pi_{j_{s-1}},v_{i_s})$ is the position immediately before $\$_{2j_{s-1}}$,  and $k=2j_{s-1}-1$ if $p_{\alpha_\dg}(\pi_{j_{s-1}},v_{i_s})$ is the position immediately after $\$_{2j_{s-1}-1}$, \\
$k'=2j_{s}$ if $p_{\alpha_\dg}(\pi_{j_{s}},v_{i_s})$ is the position immediately before $\$_{2j_{s}}$,  and $k'=2j_{s}-1$ if $p_{\alpha_\dg}(\pi_{j_{s}},v_{i_s})$ is the position immediately after $\$_{2j_{s}-1}$.
\end{quote}
\end{itemize}
\item At the same time, $\Bb$ simulates the run of $\Aa'$ over $\eta(urc_{\overline{\pi}}(\pi'))$ as follows.
\begin{itemize}
\item If $\Aa'$ makes a transition $(q,a_i,q')$ over $a_i$, then $\Bb$ checks that $b'_{i-1,0}=a_i$ and changes the state from $q$ to $q'$.
\item If $\Aa'$ makes a transition $(q,c,q')$ in the position $2i$ of $\eta(urc_{\overline{\pi}}(\pi'))$, then $\Bb$ checks that $(S_i,\chi_i,\sim_i)$ satisfies $c$, verifies that the data value in the current position is equal to the data value in the position represented by $pos_i(t)$ for each $t \in S_i$ such that $\cur \sim_i t$, and changes the state from $q$ to $q'$.
\item $\Bb$ accepts if $\Aa'$ accepts and a final profile is reached.
\end{itemize}
\end{itemize}
From the above construction, we know that in its states, $\Bb$ should record the states of $\Aa'$ and the guessed locating profiles. Because both the number of profiles and the number of functions $pos$ in locating profiles are exponential over $|\Tt_\Aa|$, it follows that the number of states of $\Bb$ is polynomial over $|Q|$ and exponential over $|\Tt_\Aa|$. 
\end{proof}

\end{appendix}

\end{document}